\documentclass{article}

\usepackage{arxiv}
\usepackage{ctable}
\usepackage[utf8]{inputenc} % allow utf-8 input
\usepackage[T1]{fontenc}    % use 8-bit T1 fonts
\usepackage{hyperref}       % hyperlinks
\usepackage{url}            % simple URL typesetting
\usepackage{booktabs}       % professional-quality tables
\usepackage{nicefrac}       % compact symbols for 1/2, etc.
\usepackage{microtype}      % microtypography
\usepackage{lipsum}		% Can be removed after putting your text content
\usepackage{graphicx}
\usepackage{natbib}
\usepackage{doi}
\usepackage{float}
\usepackage{amsfonts}%
\usepackage{amsmath}
\usepackage{subcaption} %enable the function of subfigure 
\usepackage{bm} 
\usepackage{multirow}
\usepackage{lscape}
\newcommand{\nmathbf}{\bm}

\def\bfI{{\nmathbf I}}

\def\bfX{{\nmathbf X}}
\def\bfY{{\nmathbf Y}}

\def\bft{{\nmathbf t}}

\def\bfu{{\nmathbf u}}

\def\bfx{{\nmathbf x}}
\def\bfy{{\nmathbf y}}

\def\bfgamma  {{\nmathbf \gamma}}

\def\bfepsilon{{\nmathbf \epsilon}}

\def\bfbeta   {{\nmathbf \beta}}

\def\bfLambda {{\nmathbf \Lambda}}

\def\bfSigma  {{\nmathbf \Sigma}}

\newcommand{\be}{\begin{eqnarray}}
	\newcommand{\ee}{\end{eqnarray}}
\newcommand{\ba}{\begin{eqnarray*}}
	\newcommand{\ea}{\end{eqnarray*}}

\newcommand{\reals}{\mbox{\rm I\kern-.20em R}}

\title{Bayesian hierarchical models with calibrated mixtures of g-priors for assessing treatment effect moderation in meta-analysis}

\author{ \href{https://orcid.org/0000-0002-3570-9496}{\includegraphics[scale=0.06]{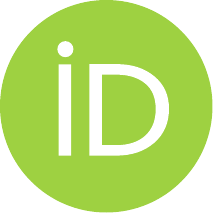}\hspace{1mm}Qiao Wang} \\
	Department of Biostatistics and Bioinformatics\\
	Duke University School of Medicine\\
	\texttt{q.wang@duke.edu} \\
	\And
	\href{https://orcid.org/0000-0002-3736-6327}{\includegraphics[scale=0.06]{orcid.pdf}\hspace{1mm}Hwanhee Hong}\thanks{Correpsonding Author} \\
	Department of Biostatistics and Bioinformatics\\
	Duke University School of Medicine\\
	\texttt{hwanhee.hong@duke.edu} \\
}

\renewcommand{\shorttitle}{\textit{arXiv} Template}

\begin{document}
	\maketitle
	
	%%% Title page
	%%% Use command\maketitle to produce the title page.
	
	\begin{abstract}
	Assessing treatment effect moderation is critical in biomedical research and many other fields, as it guides personalized intervention strategies to improve participant's outcomes. Individual participant-level data meta-analysis (IPD-MA) offers a robust framework for such assessments by leveraging data from multiple trials. However, its performance is often compromised by challenges such as high between-trial variability. Traditional Bayesian shrinkage methods have gained popularity, but are less suitable in this context, as their priors do not discern heterogeneous studies. In this paper, we propose the calibrated mixtures of g-priors methods in IPD-MA to enhance efficiency and reduce risk in the estimation of moderation effects. Our approach incorporates a trial-level sample size tuning function, and a moderator-level shrinkage parameter in the prior, offering a flexible spectrum of shrinkage levels that enables practitioners to evaluate moderator importance, from conservative to optimistic perspectives. Compared with existing Bayesian shrinkage methods, our extensive simulation studies demonstrate that the calibrated mixtures of g-priors exhibit superior performances in terms of efficiency and risk metrics, particularly under high between-trial variability, high model sparsity, weak moderation effects and correlated design matrices. We further illustrate their application in assessing effect moderators of two active treatments for major depressive disorder, using IPD from four randomized controlled trials.
	
	\end{abstract}

		%%% Main text
	%%% ------------------------------------------
	\section{Introduction}
	
Evaluation of treatment effect moderation is essential in many fields, including epidemiology, finance, and biomedical research, as it helps measure differential responses to interventions and thus informs personalized intervention strategies~\citep{kraemer2002,kraemer2013,memon2019}. However, investigating effect moderators using a single study often faces power limitations~\citep{wang2007subgroup,carly2023ss}. For example, in clinical trials, the standard recruitment strategy, which is designed to ensure sufficient power to detect the main effects, typically results in low power to identify the differential intervention effects, further reduced by the necessary corrections for multiple comparisons~\citep{berger2014subgroup}.

Individual participant data meta-analysis (IPD-MA) emerges as a robust method for synthesizing findings from all available relevant studies. By leveraging data from multiple studies, IPD-MA not only enhances statistical power, but also advances methods for evaluating the effect moderation~\citep{riley2004,hwanhee2015,burke:2017,sara2004,riley2020}. However, the application of IPD-MA in assessing effect moderation is often hampered by challenges, such as weak signals of the effect moderators and the large number of potential effect moderators~\citep{debray2015,james2013,caspar2023}. To address these challenges, several methods have been proposed to provide estimators that improve estimation risk measures, defined as the expected loss of an estimator under a given decision rule. When we adopt the quadratic loss function as a decision rule, the risk measure is the mean squared error. These methods sacrifice bias to improve efficiency (i.e., bias-variance tradeoff), mitigate estimation or prediction risk, and control model overfitting ~\citep{james2013,wee2018,sara2019,sebri2021,seo2021}.

Among the proposed methods, Bayesian shrinkage methods, such as the Bayesian least absolute shrinkage and selection operator (LASSO)~\citep{bl2008}, the horseshoe prior method~\citep{hs2010}, and the stochastic search variable selection (SSVS) method~\citep{ssvs1993}, offer significant advantages in evaluating effect moderation due to their inherent flexibility in the prior specification and their capacity for direct uncertainty quantification. Specifically, the Bayesian LASSO heavily shrinks the coefficients toward zero by placing a Laplacian prior, while the horseshoe and SSVS methods are less restrictive, allowing the coefficients that represent true effects or signals in the data to avoid excessive penalization. These methods are straightforward and easily applicable when analyzing a single study. However, they are less efficient for leveraging multiple studies in IPD-MA as they do not adequately account for varying levels of variability or shrinkage across studies.

To overcome these limitations, mixtures of g-priors offer a more flexible framework to leverage multiple studies. Mixtures of g-priors have been proposed as a Bayesian method for variable selection in s single study ~\citep{ZS1980,cui2008,Liang:2008}. In the context of multiple linear regression, $\bfY = \bfX\bfbeta + \bfepsilon,$ with $\bfY \in \reals^{n}, \bfX \in \reals^{n\times p}, \bfbeta \in \reals^{p},$ and $\bfepsilon \sim N({\bf0},\sigma^2{\bfI}_{n})$,
the g-prior for $\bfbeta$ is specified as:
\be
\label{joint:g}
\bfbeta |g, \sigma^2 \sim N_p({\bf0}, g\sigma^2(\bfX'\bfX)^{-1}).
\ee
When $g$ is prespecified or fixed, known as Zellner's g-prior, the posterior mean of $\bfbeta$ conditional on $g$ is $g/(1+g)(\bfX'\bfX)^{-1}\bfX'\bfY$. This is the product of the shrinkage factor, $g/(1+g)$, and the unbiased least squares estimator of $\bfbeta$, $(\bfX'\bfX)^{-1}\bfX'\bfY$. The shrinkage factor reflects the extent to which the posterior mean is pulled toward zero. When $g$ follows a distribution, this defines the mixtures of g-priors. In this case, the posterior mean of $\bfbeta$ is $E\left( g/(1+g)|\bfY\right)(\bfX'\bfX)^{-1}\bfX'\bfY$, where $E\left(g/(1+g)|\bfY\right)$ is the shrinkage factor, quantifying the shift in mean relative to the least squares estimator. Mixtures of g-priors offer several advantages: they provide a proper conjugate prior~\citep{bayarri2012}, establish strong theoretical foundations for model selection, produce shrinkage estimator ~\citep{zellner1962,zellner1986,som2015,qiao2023}, and enhance computational efficiency.

In this paper, inspired by the mixtures of g-priors, we propose the calibrated mixtures of g-priors methods in IPD-MA to assess treatment effect moderation by effectively leveraging multiple studies. Our approach incorporates both the trial-level and moderator-level information through the prior, aiming to improve the estimation of effect moderators with greater efficiency and reduced risk measures compared to standard Bayesian shrinkage methods. The proposed calibrated mixtures of g-priors allow different functions to integrate trial-level information and offer a wide spectrum of shrinkage levels across moderators, providing a comprehensive assessment of risk measures from conservative to optimistic approaches. 

The remainder of this paper is presented as follows. Section \ref{method} reviews widely used standard Bayesian shrinkage methods, and introduces our calibrated mixtures of g-priors methods. In Section \ref{simulation}, we evaluate the performance of our proposed methods compared to standard methods through an extensive simulation study under various settings. In Section \ref{realdata}, we apply these methods to a real data example of IPD-MA, combining four randomized controlled trials (RCTs) comparing two psychiatric treatments for patients with major depression disorder. Finally, in Section \ref{discussion}, we conclude with a discussion of the results, provide recommendations for practitioners, and suggest future research directions. 

\section{Bayesian shrinkage methods} \label{method}

\subsection{Linear IPD-MA model}
Without loss of generality, we assume that $y_{ij}$ is the continuous outcome for the $j$th participant in the $i$th trial, where $i=1,2,\cdots,I$, $j=1,2,\cdots,n_i$, and $n_i$ is the sample size for the $i$th study. The IPD-MA can then be fitted using the following linear mixed effects regression model:
\begin{align}
	y_{ij}&=\mu+t_{ij}\alpha+\bfx_{ij}\bfbeta+\bfx_{ij}^{em}\bfgamma+u_{\mu i}+t_{ij}u_{\alpha i}+\bfx_{ij}^{em}\bfu_i+\epsilon_{ij}, \label{model} \\
	u_{\mu i} &\sim N(0, \tau_{\mu}^2), u_{\alpha i} \sim N(0, \tau_{\alpha}^2), u_{ki} \sim N(0, \tau_{k}^2) \text{ for } u_{ki} \in \bfu_i,  \nonumber
\end{align}
where $\mu$ is the intercept, $\alpha$ is the conditional treatment effect with the treatment indicator $t_{ij}$ (1 for the treatment and 0 for the control group), $\bfbeta=(\beta_1, \cdots, \beta_p) \in \reals^{p}$ is a set of coefficients for the centered baseline covariates $\bfx_{ij}^{\prime}=(x_{ij1},\cdots,x_{ijp})^{\prime} \in \reals^{p}$, $\bfgamma=(\gamma_1, \cdots, \gamma_d) \in \reals^{d}$ represents the coefficients for the interactions between the treatment and potential effect moderators $\bfx_{ij}^{em \prime}=(t_{ij}x_{ij1}, \cdots, t_{ij}x_{ijd})^{\prime}\in \reals^{d}$, where $0 \leq d \leq p$, $u_{\mu_i}$ and $u_{\alpha_i}$ are the random effects for the intercept and the conditional treatment effect with $\tau_\mu^2 \mbox{ and } \tau_\alpha^2 $ capturing the between-trial variability respectively. $\bfu_{i}=(u_{1i}, \cdots, u_{ki} \cdots,u_{di}) \in \reals^{d}$ is a vector of random effects for the potential effect moderators with covariates $\bfx_{ij}^{em} $, where $u_{ki}$ models the random effects of the $k^{th}$ effect moderator with $\tau_k^2$ capturing its between-trial variability. Finally, $\epsilon_{ij} \sim N(0,\sigma_i^2)$ is the random noise for study $i$. Although it is also feasible to specify random effects for the baseline effects $\bfbeta$, we consider them fixed for simplicity. Here, the parameter of interest is the magnitude of the moderation effect, $\bfgamma$.  

As our main focus is on assessing effect moderators, we adopt a non-informative prior, $\pi(\mu,\alpha,\bfbeta_,\sigma^2) \propto 1/\sigma^2$ for coefficients other than $\bfgamma$. For the parameters of variability between trials, we use $\tau_\mu,\tau_\alpha, \tau_k \sim  C^{+}(0,1)$, where $C^{+}(0,1)$ is a half Cauchy distribution~\citep{gelman2006}. Exploring the optimal prior distributions for $\tau_\mu,\tau_\alpha, \mbox{and } \tau_k$ falls outside the scope of this paper; however, and we anticipate that varying these priors will have minimal impact on the performance comparison of our proposed methods. 

In the following subsections, we review existing Bayesian shrinkage methods, and introduce our calibrated mixtures of g-priors methods. Table \ref{summary:methods} summarizes all Bayesian methods considered in this paper.

\subsection{Two existing Bayesian shrinkage methods}

\subsubsection{Horseshoe prior}

The horseshoe (HS) prior~\citep{hs2010} is often used for variable selection with high-dimensional data. This prior differentiates the global and local shrinkage, facilitating the Bayesian estimation in a sparse model, and is defined as follows: 
\begin{equation}
	\gamma_k|\lambda_k,\tau \sim N(0,\lambda_k^2\tau^2),~\lambda_k \sim C^{+} (0,1),~ \tau \sim C^{+}(0,1), \nonumber
\end{equation}
where $\lambda_k$ governs local shrinkage and $\tau$ controls global shrinkage. The horseshoe prior was further extended as the regularized horseshoe prior to incorporate the degree of sparsity and control the shrinkage for large coefficients through hyperparameters ~\citep{piironen2017}, but we consider the basic horseshoe prior.

\subsubsection{Stochastic search variable selection}

The stochastic search variable selection (SSVS) method is another popular Bayesian method for model selection in high-dimensional data~\citep{ssvs1993,ishwaran2005,hara2009,seo2021}. Unlike the horseshoe prior, which utilizes scale mixtures, SSVS comprises a mixture of two normal distributions. Although numerous variants of SSVS have been proposed for various applications, 

we employ the SSVS specified in~\cite{seo2021} as follows: 
\be
\gamma_k|I_k,h_k,\eta, \sim (1-I_k)N(0,\eta^2)+I_k N(0,h_k\eta^2). \nonumber
\ee
The first component controls shrinkage by centering a point mass around zero with $\eta \sim Unif(0,c)$, where $c$ is a small constant such as 5. The second component lessens the shrinkage for the true effect moderator with a large $h_k$, such as 100. Finally, $I_k \sim Ber(0.5)$ serves as a binary indicator of the presence of a true effect moderator.

\subsection{Na\"ive mixtures of g-priors in IPD-MA}  \label{ma:g-prior:section}
In model \eqref{model}, the simple mixtures of g-priors in \eqref{joint:g} is respecified as follows:
\begin{equation*}
	% \label{joint:g:false}
	\bfgamma |g, \sigma^2 \sim N_d\left({\bf0}, g\sigma^2(\bfX^{em'}\bfX^{em})^{-1}\right), g>0,
\end{equation*}
where $\bfX^{em}$ is the design matrix with the $(i,j)$ element denoted by $\bfx^{em}_{ij}$. The simple g-prior has several limitations. First, it imposes uniform shrinkage across both true and false effect moderators via the parameter $g$, leading to excessive shrinkage of true effect moderators, while insufficiently shrinking coefficients of the non-effect moderators. Second, the assumption that the noise parameters are the same in all studies (i.e., $\sigma_i$=$\sigma$ for any $i$) is overly stringent. Third, the requirement that $\bfX^{em}$ has full column rank in order for $(\bfX^{em'}\bfX^{em})^{-1}$ to exist is overly restrictive, particularly when dealing with a large number of potential effect moderators that are highly correlated. 

To address these issues, we first propose the na\"ive mixtures of g-priors (NMG) methods for $\gamma_k$ in model \eqref{model} as follows:  
\be
\label{ma:gprior}
\gamma_k | g_k,\bft,\bfx_{[k]},\bfLambda \sim N\left(0,g_{k}[(\bft\circ \bfx_{[k]})'\bfLambda(\bft\circ \bfx_{[k]})]^{-1}\right), g_k>0,
\ee
where $g_k$ is the shrinkage parameter for the $k^{th}$ potential effect moderator, $\circ$ is the Hadamard product, $\bft=(t_{11},\cdots,t_{n_1},\cdots,t_{I1},\cdots,t_{In_{I}})'$ is the vector of treatment indicators for all participants across $I$ studies, $\bfx_{[k]}=(x_{11k},\cdots,x_{1{n_1}k},\cdots,x_{I1k},\cdots,x_{In_{I}k})'$ is the vector of covariates for the $k^{th}$ effect moderator,  $\bfLambda=diag(1/\sigma_1^2{\bf1}_{n_1},\cdots,1/\sigma_I^2{\bf1}_{n_{I}})$ scaling $\bft\circ \bfx_{[k]}$ according to each study's specific $\sigma_i^2$. The prior in \eqref{ma:gprior} for a single study is commonly referred to as ``independent mixtures of g-priors'' or ``block mixtures of g-priors'' ~\citep{ley2012,som:2016,MS2016}. We consider several specifications for $g_k$ in the NMG methods.

\subsubsection{Unit information prior in NMG}

The unit information prior (UIP) prespecifies $g_k$ based on the total sample size of all studies:
\begin{equation*}
	g_k=N=\sum_{i=1}^{I}n_i.
\end{equation*}
This represents the prior's contribution equivalent to one observation's worth of information~\citep{kw1995,Liang:2008}.

\subsubsection{Zellner-Siow prior in NMG}
The Zellner-Siow (ZS) prior~\citep{ZS1980} utilizes an inverse-Gamma (IG) distribution for $g_k$, incorporating the total sample size into its scale parameter. The ZS prior in IPD-MA is defined as:  
\begin{equation*}
	\label{ma:zs}
	g_{k}|N \sim IG \left(1/2, N/2\right).
\end{equation*}
Integrating $g_k$ from \eqref{ma:gprior} results in $\gamma_k|\bft,\bfx_{[k]},\bfLambda$ following a Cauchy distribution. Therefore, the ZS prior together with \eqref{ma:gprior} is the hierarchical representation of $\gamma_k|\bft,\bfx_{[k]},\bfLambda$, providing computational efficiency.

\subsubsection{Hyper-g prior in NMG}
The hyper-g (HG) prior, proposed by~\cite{Liang:2008}, specifies a hyperprior to $g_k$ as follows: 
\begin{equation*}
	\label{hyperg}
	\pi(g_k|a)=\frac{(a-2)}{2}(1+g_k)^{-a/2}.
\end{equation*}
This implies that $g_k/(1+g_k) \sim Beta(1,a/2-1)$, where $Beta(r,s)$ denotes a Beta distribution with shape parameter $r$ and scale parameter $s$. Different values of $a$ lead to various prior shrinkage levels for $\gamma_k$. Commonly used values for $a$ are 3 and 4; $a=3$ places more mass of the shrinkage factor $g_k/(1+g_k)$ around 1, while $a=4$ results in a uniform prior to the shrinkage factor. 

\subsubsection{Hyper-g/N prior in NMG}
The hyper-g/N (HGN) prior, proposed by \cite{Liang:2008}, is an extended version of the HG prior. It adapts the level of prior influence based on the number of observations, and it is specified as:  
\begin{equation*}
	\label{ma:hgn}
	\pi(g_k|a,N)=\frac{(a-2)}{2N}(1+\frac{g_k}{N})^{-a/2},
\end{equation*}
which implies that $g_k/(g_k+N) \sim Beta(1,a/2-1)$. Incorporating the sample size $N$ in the formula addresses the model selection inconsistency observed in the HG prior. That is, the HG prior fails to guarantee posterior probabilities converging to 1 for the true model as the sample size increases. Similarly, we consider $a=3$ and $a=4$ for the HGN prior. %.

\subsubsection{Limitations of NMG}
The NMG methods offer a straightforward framework for applying mixtures of g-priors when combining multiple studies in IPD-MA. However, they still present two key limitations. First, the HG prior applies the $Beta(1,a/2-1)$ distribution to $g_k/(1+g_k)$, where the distribution increases monotonically for $2 < a < 4$ and decreases for $a > 4$. Although choosing $2 < a < 4$ ensures a proper prior and mitigates excessive shrinkage, this monotonicity causes the shrinkage factor's distribution to be skewed heavily towards 1 or 0. While this may be beneficial for model selection, it is less suitable for parameter estimation. To address this, a heavy-tailed hyperprior distribution for $g_k/(1+g_k)$ needs to be studied to allow for more flexible shrinkage. 

Second, the UIP, ZS, and HGN priors rely on the total number of observations $N$ from all studies in the IPD-MA. While including $N$ in the Bayesian model selection is common with two typical examples of the Bayesian Information Criterion (BIC) and the ZS prior, \cite{berger2014} and \cite{bayarri2019} have proposed to use the effective sample size (TESS) instead of a simple number of observations to achieve model selection consistency. Drawing inspirations from the TESS, it becomes apparent that using $N$ in the prior assumes that each participant contributes equally to the estimation of $\gamma_k$, overlooking differences in individuals across studies and imposing insufficient shrinkage of the estimates. Thus, alternative approaches that better leverage the total number of observations in the hyperprior for $\gamma_k$ are needed to improve the estimation of moderation effects. However, the formal derivation of TESS is designed for the model selection and remains complex even with a single study, we adopt a more data-driven approach, described in the following section, to calibrate the sample size contribution for each study. 

\subsection{Calibrated mixtures of g-priors in IPD-MA} \label{ma:cmg:section}

To address the aforementioned limitations of the NMG methods, we propose the calibrated mixtures of g-priors (CMG) methods in IPD-MA as follows:
\begin{align}
	\gamma_k | g_k,\bfLambda^{\star},\bft,\bfx_{[k]} &\sim N\left(0,g_{k}[(\bft\circ \bfx_{[k]})'\bfLambda^{\star}(\bft\circ \bfx_{[k]})]^{-1}\right), g_k>0, \label{calib:gamma} \\ 
	\bfLambda^{\star} &= diag\left(\sigma_1^{-2}f(n_1)^{-1}{\bf1}_{n_1},\cdots,\sigma_I^{-2}f(n_I)^{-1}{\bf1}_{n_{I}}\right), \label{calib:g} \nonumber 
\end{align}
where $f(n_i)$ in $\bfLambda^{\star}$ is the study-specific sample size tuning function, which scales the design matrix within each study. The CMG methods adjust different shrinkage levels across potential effect moderators through $g_k$, and control the expected contribution of observations to the estimation within each study through $f(n_i)$. We consider 9 different prior specifications for the CMG methods, representing combinations of three hyperpriors for $g_k$ and three functions for $f(n_i)$ (see Table \ref{summary:methods}).

\subsubsection{Hyperprior for $g_k$ in CMG}
\label{hyperprior:g}

Motivated by the HG prior, three hyperpriors for $g_k$ to reflect  different levels of prior shrinkage are defined as follows:
\begin{itemize}
	\item Shrinkage 1 (S1):
	\begin{equation*}
		\label{ma.cmg.s1}
		\pi_{S_1}(g_k|b_k) = \frac{g_k}{(1+g_k)^{b_k+2}B(2,b_k)}, b_k \leq 2,
	\end{equation*}
	where $B(2,b_k)$ is the Beta function. The S1 hyperprior is equivalent to  $g_k/(1+g_k) \sim Beta(2,b_k)$. Here, $b_k$ is used to control the skewness of $g_k/(1+g_k)$, with $b_k$ either prespecified as a value less than 2 or assigned a hyperprior. For example, we assume $b_k \sim Uniform (0,2)$, then the S1 hyperprior represents the least shrinkage, with a conceptually expected prior shrinkage of $E(g_k/(1+g_k))=ln(2) \approx 0.69$, indicating an average prior shrinkage of 31\% toward 0 in coefficients. 
	
	\item Shrinkage 2 (S2):
	\begin{equation*}
		\label{ma.cmg.s2}
		\pi_{S_2}(g_k) = \frac{1}{(1+g_k)^2},
	\end{equation*}
	which is equivalent to $g_k/(1+g_k) \sim Beta(1,1)$, implying a conceptually average prior shrinkage of 50\% toward 0 in coefficients. This is the same as the HG prior in \eqref{hyperg} with $a=4$ and $f(n_i)=1$.
	\item Shrinkage 3 (S3):
	\begin{equation*}
		\label{ma.cmg.s3}
		\pi_{S_3}(g_k|b_k) = \frac{g_k^{b_k-1}}{(1+g_k)^{b_k+2}B(b_k,2)}, b_k \leq 2,
	\end{equation*}
	which is equivalent to $g_k/(1+g_k) \sim Beta(b_k,2)$, where $b_k \leq 2$ ensures that $g_k/(1+g_k)$ is right-skewed. We follow the same rule as in S1 to specify $b_k$. If $b_k \sim Uniform(0,2)$, S3 results in a conceptually expected prior shrinkage of 69\% toward 0 in coefficients. 
\end{itemize}

In summary, the S1 hyperprior results in the least shrinkage of point estimates of $\gamma_k$ toward 0, while S3 produces the most shrinkage when the same $f(n_i)$ is used. When S3 is applied to all potential effect moderation, the resulting point estimates of $\gamma_k$ are the most conservative (i.e., close to the null value). This suggests that an effect of moderator with a relatively large estimated effect size under S3 may have a stronger influence compared to those with small effect sizes. We denote the CMG methods under each shrinkage level with the suffixes -S1, -S2, and -S3. Figures A1, A2, and A3 in the supplementary material visualizes marginal density of $g_k$, marginal density of $g_k/(1+g_k)$, and conditional density of $g_k/(1+g_k)$ under three specifications, respectively.

\subsubsection{Hyperprior for $f(n_i)$ in CMG} 
\label{hyperprior:f}

The sample size tuning function $f(n_i)$ for study $i$ in the CMG methods is a function of the number of observations $n_i$. It is designed to calibrate how many observations are expected to contribute to the estimation of $\gamma_k$. Additionally, we impose the constraint $f(n_i) \leq n_i$, ensuring that the contribution does not exceed the total number of observations in each study. This constraint can be implemented by including an unknown tuning parameter $p_i$ in $f(n_i)$, and we consider three functions:
\begin{itemize}
	\item $n$: $f(n_i|p_i)=n_ip_i,p_i \sim Uniform(1/n_i,1)$; 
	\item $log$: $f(n_i|p_i)=log(n_ip_i),p_i \sim Uniform(1/n_i,1)$;
	\item $pow$: $f(n_i|p_i)=n_i^{p_i},p_i \sim Uniform(0,1)$.
\end{itemize}
Here, $p_i$ controls the number of observations expected to be leveraged, with at least one observation and up to $n_i$ observations. By integrating $p_i$ out of $f(n_i|p_i)$, the ``$log$'' function has the slowest rate of increase, while the ``$n$'' function exhibits the fastest increase. We denote the CMG methods under each of the trial-level tuning functions with the suffixes -n, -log, and -pow. Figure A4 in the supplementary material illustrates different rates of increase these three sample size tuning functions as the sample size varies, with a fixed tuning parameter $p_i$.

\subsection{Other specifications for CMG}
In addition to the hyperpriors introduced in Section \ref{hyperprior:g}, other forms of $g_k$ and $f(n_i)$ can be considered in the CMG methods. In this subsection, we introduce two additional forms of the CMG methods.

\subsubsection{Calibrated UIP}

The calibrated UIP (CUIP) is defined by setting $g_k=1$ and $f(n_i)=\sqrt{N}$ (or equivalently, $g_k=\sqrt{N}$ and $f(n_i)=1$) in the CMG methods. This simplifies the trial-level sample size calibration to an overall sample size adjustment, reducing the number of parameters and providing a simple calibration. A similar specification was studied in analyzing a single study~\citep{pans2022}.

\subsubsection{Calibrated ZS prior}

The extension of the ZS prior to the calibrated ZS prior (CZS) requires the representation of the ZS prior. Note that the ZS prior in \eqref{ma:zs} is equivalent to:
\begin{equation*}
	\gamma_k | g_k,N,\bfLambda,\bft,\bfx_{[k]} \sim N\left(0,g_{k}N[(\bft\circ \bfx_{[k]})'\bfLambda(\bft\circ \bfx_{[k]})]^{-1}\right), g_k \sim IG(1/2,1/2).  
\end{equation*}
The CZS is then defined by replacing $\bfLambda/N$ with $\bfLambda^{\star}$, and applying $g_k \sim IG(1/2,1/2)$. For CZS, we only consider the sampling size tuning function $f(n_i|p_i)=n_ip_i$ in $\bfLambda^{\star}$ with $p_i \sim Uniform(1/n_i,1)$ for simplicity.  

\subsection{Computation}  

Our proposed methods can be efficiently implemented using Markov Chain Monte Carlo (MCMC) via Just Another Gibbs Sampler (JAGS) within the R environment. For both simulation studies and data analyses, we use two MCMC chains, each with 20,000 samples. To improve the chain mixing properties, we apply thinning by retaining every $10^{th}$ sample and discard the first 10,000 samples as burn-in. A sample R code for the methods listed in Table \ref{summary:methods} is available from \textit{https://github.com/QWCodeShare/ModerationPaper.git}. 

\begin{landscape}
	\begin{table}[t]
		\caption{Summary of Methods}
		\vspace{2em}
		\hspace{-0.5em}
		\label{summary:methods}
		\resizebox{1.0\linewidth}{!}{%
			\begin{tabular}{lll}
				\specialrule{.1em}{.05em}{.05em}
				\textbf{Methods} &
				\textbf{Prior distribution for $\gamma_k$} &
				\textbf{Hyperparameter specification} \\ \hline
				\textbf{Non-shrinkage Method} & &  \\   
				Flat &
				$\gamma_k \propto 1$ &
				\\ \hline
				\textbf{Two Existing Shrinkage Methods} & &  \\   
				HS &
				$\gamma_k|\lambda_k,\tau \sim N(0,\lambda_k^2\tau^2)$ &
				$\lambda_k \sim C^{+} (0,1), \tau \sim C^{+}(0,1)$ \\
				SSVS &
				$\gamma_k|I_k,h_k,\eta, \sim (1-I_k)N(0,\eta^2)+I_k N(0,h_k\eta^2)$ &
				$I_k \sim Ber(0.5), \eta \sim Uniform(0,c_1)$ with small $c_1$ \\ \hline
				\textbf{NMG Methods} & &  \\ 
				UIP &
				$\gamma_k | g_k,\bfLambda \sim N\left(0,g_{k}(\bfx_{k}'\bfLambda\bfx_{k})^{-1}\right)$ &
				$g_k=N$ \\
				ZS &
				$\gamma_k | g_k,\bfLambda \sim N\left(0,g_{k}(\bfx_{k}'\bfLambda\bfx_{k})^{-1}\right)$ &
				$g_{k}|N \sim IG \left(1/2, N/2\right)$ \\
				
				HG $(a=3)$ &
				$\gamma_k | g_k,\bfLambda \sim N\left(0,g_{k}(\bfx_{k}'\bfLambda\bfx_{k})^{-1}\right)$ &
				$g_k/(g_k+1) \sim Beta(1,3/2-1)$ \\
				HG $(a=4)$ &
				$\gamma_k | g_k,\bfLambda \sim N\left(0,g_{k}(\bfx_{k}'\bfLambda\bfx_{k})^{-1}\right)$ &
				$g_k/(g_k+1) \sim Beta(1,4/2-1)$ \\
				HGN $(a=3)$ &
				$\gamma_k | g_k,\bfLambda \sim N\left(0,g_{k}(\bfx_{k}'\bfLambda\bfx_{k})^{-1}\right)$ &
				$g_k/(g_k+N) \sim Beta(1,3/2-1)$ \\
				HGN $(a=4)$ &
				$\gamma_k | g_k,\bfLambda \sim N\left(0,g_{k}(\bfx_{k}'\bfLambda\bfx_{k})^{-1}\right)$ &
				$g_k/(g_k+N) \sim Beta(1,4/2-1)$ \\ \hline
				
				\textbf{CMG Methods} & &  \\   
				CUIP &
				$\gamma_k | g_k,\bfLambda \sim N\left(0,g_{k}(\bfx_{k}'\bfLambda^{\star}\bfx_{k})^{-1}\right)$ &
				$g_k=1$, $f(n_i)=\sqrt{N}$ \\
				
				CZS &
				$\gamma_k|g_k,\bfLambda^{\star}  \sim N\left(0,g_{k}(\bfx_{k}'\bfLambda^{\star}\bfx_{k})^{-1}\right)$ &
				$g_{k} \sim IG \left(1/2, 1/2\right),~ f(n_i)=n_ip_i, ~ p_i \sim Uniform(1/n_i,1)$ \\
				
				CMG-S1-n &
				$\gamma_k|g_k,\bfLambda^{\star}  \sim N\left(0,g_{k}(\bfx_{k}'\bfLambda^{\star}\bfx_{k})^{-1}\right)$ &
				$g_k/(g_k+1) \sim Beta (2,b_i),~f(n_i)=n_ip_i, ~ b_i \sim Uniform(0,2),~ p_i \sim Uniform(1/n_i,1)$ \\
				CMG-S2-n &
				$\gamma_k|g_k,\bfLambda^{\star}  \sim N\left(0,g_{k}(\bfx_{k}'\bfLambda^{\star}\bfx_{k})^{-1}\right)$ &
				$g_k/(g_k+1) \sim Beta (1,1),~f(n_i)=n_ip_i,  ~ b_i \sim Uniform(0,2),~ p_i \sim Uniform(1/n_i,1)$ \\
				CMG-S3-n &
				$\gamma_k|g_k,\bfLambda^{\star}  \sim N\left(0,g_{k}(\bfx_{k}'\bfLambda^{\star}\bfx_{k})^{-1}\right)$ &
				$g_k/(g_k+1) \sim Beta (b_i,2),~f(n_i)=n_ip_i,  ~ b_i \sim Uniform(0,2),~ p_i \sim Uniform(1/n_i,1)$ \\
				CMG-S1-log &
				$\gamma_k|g_k,\bfLambda^{\star}  \sim N\left(0,g_{k}(\bfx_{k}'\bfLambda^{\star}\bfx_{k})^{-1}\right)$ &
				$g_k/(g_k+1) \sim Beta(1,1),~f(n_i)=log(n_ip_i),  ~ b_i \sim Uniform(0,2),~ p_i \sim Uniform(1/n_i,1)$ \\
				CMG-S2-log &
				$\gamma_k|g_k,\bfLambda^{\star}  \sim N\left(0,g_{k}(\bfx_{k}'\bfLambda^{\star}\bfx_{k})^{-1}\right)$ &
				$g_k/(g_k+1) \sim Beta (b_i,2),~f(n_i)=log(n_ip_i),  ~ b_i \sim Uniform(0,2),~ p_i \sim Uniform(1/n_i,1)$ \\
				CMG-S3-log &
				$\gamma_k|g_k,\bfLambda^{\star}  \sim N\left(0,g_{k}(\bfx_{k}'\bfLambda^{\star}\bfx_{k})^{-1}\right)$ &
				$g_k/(g_k+1) \sim Beta (2,b_i),~f(n_i)=log(n_ip_i),  ~ b_i \sim Uniform(0,2),~ p_i \sim Uniform(1/n_i,1)$ \\
				CMG-S1-pow &
				$\gamma_k|g_k,\bfLambda^{\star}  \sim N\left(0,g_{k}(\bfx_{k}'\bfLambda^{\star}\bfx_{k})^{-1}\right)$ &
				$g_k/(g_k+1) \sim Beta (2,b_i),~f(n_i)=n_i^{p_i}, ~ b_i \sim Uniform(0,2),~ p_i \sim Uniform(0,1)$ \\
				CMG-S2-pow &
				$\gamma_k|g_k,\bfLambda^{\star}  \sim N\left(0,g_{k}(\bfx_{k}'\bfLambda^{\star}\bfx_{k})^{-1}\right)$ &
				$g_k/(g_k+1) \sim Beta (1,1),~f(n_i)=n_i^{p_i}, ~ b_i \sim Uniform(0,2),~ p_i \sim Uniform(0,1)$ \\
				CMG-S3-pow &
				$\gamma_k|g_k,\bfLambda^{\star}  \sim N\left(0,g_{k}(\bfx_{k}'\bfLambda^{\star}\bfx_{k})^{-1}\right)$ &
				$g_k/(g_k+1) \sim Beta (b_i,2),~f(n_i)=n_i^{p_i}, ~ b_i \sim Uniform(0,2), ~ p_i \sim Uniform(0,1)$ \\ \hline
				\specialrule{.1em}{.05em}{.05em}
		\end{tabular}}
	\end{table}
\end{landscape}

\section{Simulation} \label{simulation}
Our simulation study has three objectives. First, we assess and compare the performance of 20 different methods listed in Table \ref{summary:methods} in estimating moderation effects. Second, we study the bias-variance tradeoff and participant-specific estimation across all methods. Third, we discern specific settings where our proposed CMG methods offer better performance than other methods. 

\subsection{Simulation setup}
\label{sim:setup}

We generate IPD-MA data containing five trials ($I=5$) of which the sample size $n_i$ varies from 100 to 150. We consider 8 baseline covariates (i.e.,$p=8$) of which a subset is the true effect moderators. Continuous outcomes are simulated based on the model \eqref{model} with true parameters set to an intercept of $\mu=2$, a conditional treatment effect of $\alpha=3$, coefficients for baseline covariates of $\bfbeta=(1.8,2.7,2.3,1.5,1.7,2.2,1.3,2.6)$, noise $(\sigma_1,\cdots,\sigma_5)=(3.5,2.5,2.1,2.8,3)$, between-trial (i.e., between-study) variability for the intercept of $\tau_{\mu} =1.5$, between-trial variability for the conditional treatment effect of $\tau_\alpha=1.5$. 

We vary true value settings for the remaining parameters of moderation effects, and Table A1 in the supplementary material summarizes all settings. Hereafter, in all tables and figures, we use ``EM'' as the abbreviation of effect moderation to save space. 
\begin{itemize}
	\item[(a)] We consider three settings for the between-trial variability of the moderation effect $\gamma_k$. Different between-trial variability for the $k^{th}$ effect moderator is set through $\tau_k$: $\tau_k \in (1.5,2.5)$ represents high variability (i.e., $\tau_k$ varies from 1.5 to 2.5), $\tau_k \in (0.5,1.5)$ is medium variability, and $\tau_k =0$ means no variability. 
	\item[(b)] We consider three settings for the model sparsity. The model sparsity is determined by the number of true effect moderators of which the corresponding coefficients elements in $\bfgamma$ are non-zero. High sparsity corresponds to 2 out of 8 covariates being true effect moderators.Medium sparsity sets 4 out of 8 covariates as true effect moderators, and low sparsity sets 6 out of 8 covariates as true effect moderators.
	\item[(c)] We consider two settings for the correlation matrices when simulating baseline covariates $\bfx_{ij}$ from a multivariate normal distribution (MVN) with $\bfx_{ij} {\sim} MVN({\bf0}, \bfSigma_i^{B})$. Here, $\bfSigma_{i}^{B} = \sigma_{is}^{B}\sigma_{it}^{B}\rho_{ist}^{B}$, with $\sigma_{is}^{B}=\sigma_{it}^{B}=1$ representing the standard deviations of baseline covariates $s$ and $t$, and $\rho_{ist}^{B}$ representing the correlation among these characteristics. We consider no correlation (i.e., $\rho_{ist}^{B}=0$), and high correlation $0.5<\rho_{ist}^{B}<0.9$ for $s \neq t$ among baseline covariates.
	\item[(d)] We vary the strength of the true effect moderators. The first setting is strong effect moderators with $\gamma_k = 1.5$ (i.e., 50\% of the size of the conditional treatment effect, $\alpha=3$). 
	The second setting is weak effect moderators with $\gamma_k = 0.75$ (i.e., 25\% of the size of the conditional treatment effect, $\alpha=3$). 
\end{itemize}
As a result, we consider a total of 36 simulation scenarios. For each scenario, we simulate 500 datasets, and fit the 20 methods listed in Table \ref{summary:methods}.

\subsection{Simulation performance assessment}
The relative performances of methods are evaluated by three metrics. First, for $\bfgamma \in \reals^{d}$ (i.e., $d=8$), we calculate the average relative root mean squared error (ARRMSE) defined as: 
\begin{align}	
	\label{def:arrmse}
	ARRMSE(\bfgamma)=& \dfrac{\sqrt{\sum_{r=1}^{500}\sum_{k=1}^{d}(\hat{\gamma}_{k}^{(r)}-\gamma_{k})^2}}{500 d\sum_{k=1}^{d}|\gamma_{k}|}.
\end{align}
Similarly, we calculate the average absolute relative bias (AARBias):
\begin{align}	
	AARBias(\bfgamma)=& \dfrac{\sqrt{\sum_{r=1}^{500}\sum_{k=1}^{d}|\hat{\gamma}_{k}^{(r)}-\gamma_{k}|}}{500 d\sum_{k=1}^{d}|\gamma_{k}|},
\end{align}
and the average relative standard deviation (ARSD):
\begin{align}
	ARSD(\bfgamma)=& \dfrac{\sqrt{\sum_{r=1}^{500}\sum_{k=1}^{d}\left(\hat{\gamma}_{k}^{(r)} - \bar{\gamma}_{k}\right)^2}}{500 d\sum_{k=1}^{d}|\gamma_{k}|},
\end{align}
where $\bar{\gamma}_{k} = \sum_{r=1}^{500} \hat{\gamma}_{k}^{(r)}/500$. 
Finally, to investigate how the estimation of coefficients behave after taking into account the participant-level information, we calculate the participant-specific root mean squared error (PSRMSE):
\begin{align}
	\label{psrmse}
	PSRMSE= \sqrt{||(t_{ij}\alpha+\bfx_{ij}^{em}\bfgamma)-(t_{ij}\hat{\alpha}+\bfx_{ij}^{em}\hat{\bfgamma})||/N},
\end{align}
where ``||$\cdot$||'' is the Euclidean norm. The participant-specific root mean squared error for effect moderation (PSRMSE(EM)) is defined as:
\begin{align}
	\label{psrmse:em}
	PSRMSE(EM)= \sqrt{||\bfx_{ij}^{em}\bfgamma-\bfx_{ij}^{em}\hat{\bfgamma}||/N}.
\end{align}

ARRMSE, AARBias and ARSD are scalar metrics free from unit and dimension to assess the estimations of treatment effect moderation, with smaller values indicating better performance. In contrast, PSRMSE and PSRMSE(EM) are not free from unit and dimension, but reflect the estimation risk at a participant-level by taking into account the participant's information via $\bfx_{ij}$. We expect that shrinkage methods will yield smaller values of the risk metrics including ARRMSE, PSRMSE and PSRMSE(EM), particularly in challenging cases such as high between-trial variability or high model sparsity. In addition, we expect that the CMG methods result in better performance than other methods. 

\subsection{Simulation results} \label{results}

\begin{figure}[htbp]
	\centering
	\includegraphics[scale=0.7]{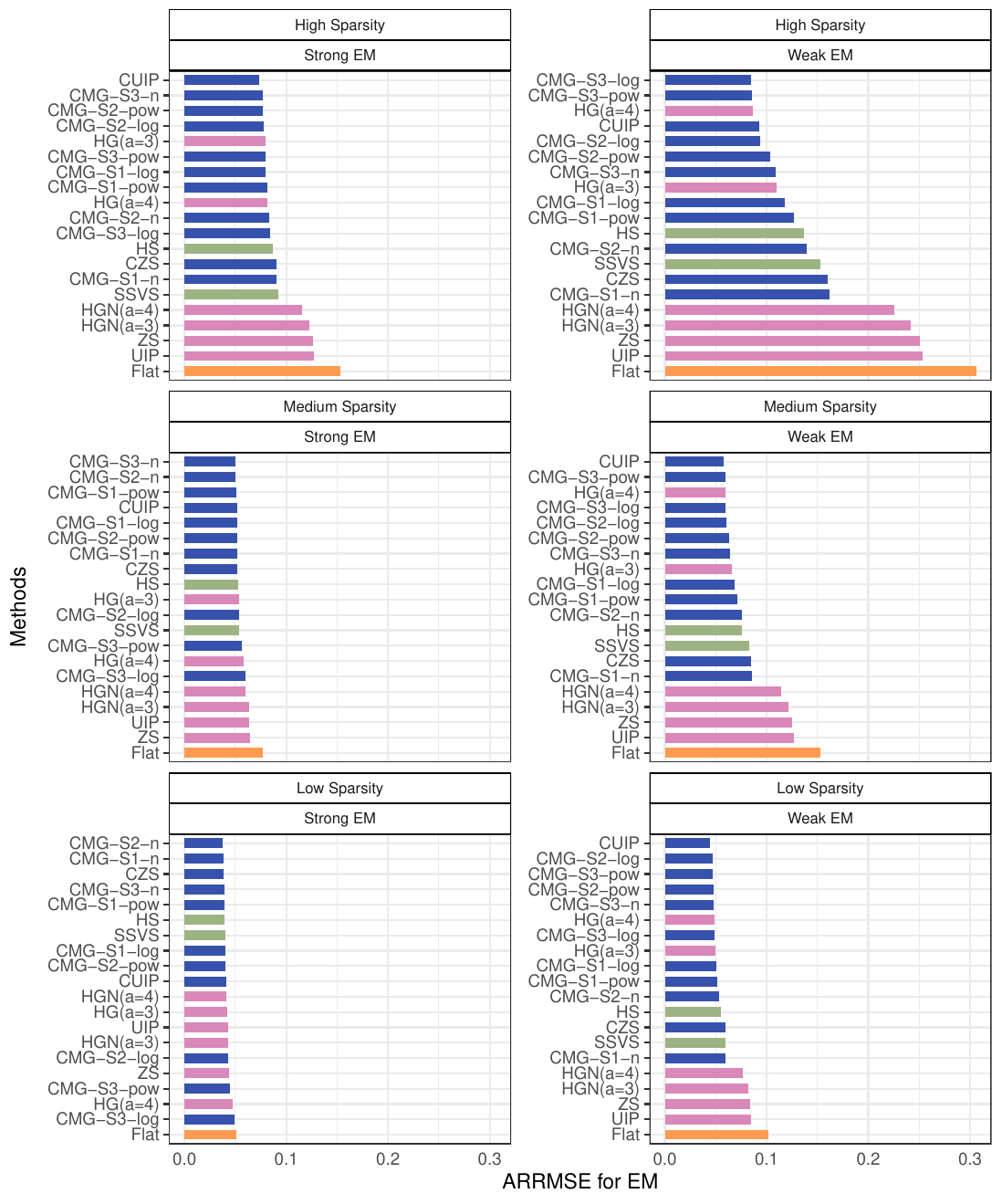}
	\caption{ARRMSE of moderation effects ($\bfgamma$) across 20 methods under a combination of three model sparsity settings (from top to bottom panels) and two magnitude of effect moderation settings (left and right panels) with high between-trial variability and correlated covariates. In each panel, methods are ordered by their performance. The blue, pink, green, and orange bars correspond to the CMG, NMG, HS and SSVS, and Flat methods, respectively.}
	\label{simu:ht}
\end{figure}

Figure \ref{simu:ht} shows how ARRMSE varies across methods under a combination of three model sparsity settings (from top to bottom panels) and two magnitude of effect moderation settings (left and right panels) with high between-trial variability and correlated baseline covariates. Overall, the CUIP, CMG-S3-n, CMG-S2-pow, CMG-S1-log, CMG-S1-pow, and CMG-S2-n methods are ranked among top 10 methods across all six scenarios with small ARRMSE, indicating that these methods are less sensitive to misspecification of the true magnitude of moderator effect and model sparsity. In contrast, the HGN (a=3), HGN (a=4), ZS, UIP and Flat methods are ranked bottom 10 with large ARRMSE, and the Flat method uniformly performs the worst across all cases. This highlights the need of calibration for NMG methods. Furthermore, as model sparsity increases (from top to bottom panels) or as the magnitude of the moderator decreases (from left to right panels), the CMG methods tend to produce smaller ARRMSE values, and the differences among the methods become more pronounced. In specific, the CMG methods perform better than other methods with respect to ARRMSE under the setting of high model sparsity and small moderation effect (see bottom right panel). In this setting, among the CMG methods (results shown with blue bars in each figure), the CMG-S3-log method provides the smallest ARRMSE, corresponding to the maximum level of prior shrinkage and closely reflecting the true data generation mechanism. 

\begin{figure}[htpb]
	\hspace{-3em}
	\centering			
	\includegraphics[scale=0.72]{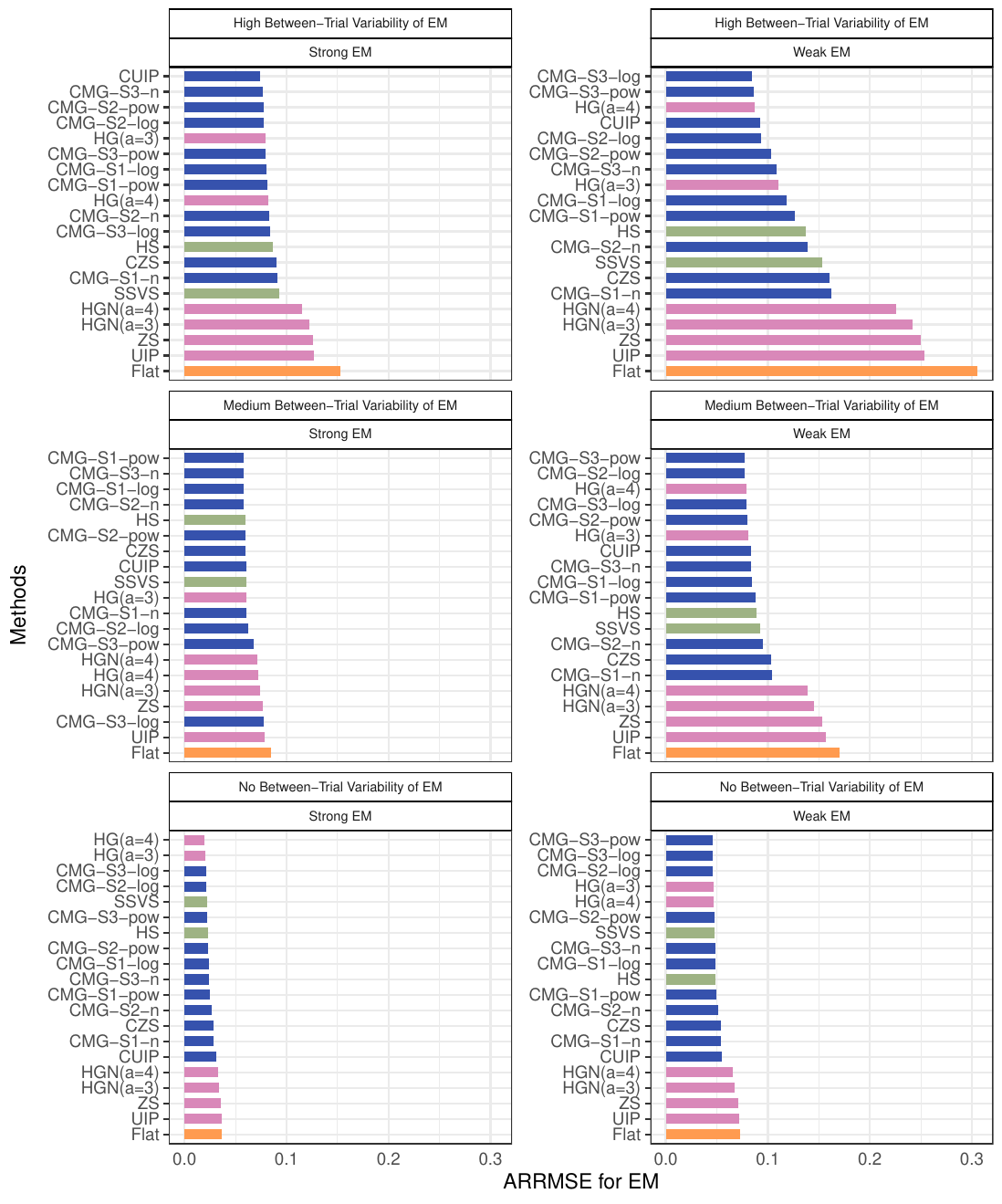}
	\caption{ARRMSE of moderation effects ($\bfgamma$) across 20 methods under the combination of three between-trial variability settings (from top to bottom panels) and two magnitude of effect moderation settings (left and right panels) with high model sparsity and correlated covariates. In each panel, methods are ordered by their performance. The blue, pink, green, and orange bars correspond to the CMG, NMG, HS and SSVS, and Flat methods, respectively.}
	\label{simu:pc}
\end{figure}

Figure \ref{simu:pc} presents how ARRMSE changes across methods under a combination of three between-trial variability settings (from top to bottom panels) and two magnitudes of effect moderation settings (left and right panels) with high model sparsity and correlated covariates. Overall, the CMG-S3-n, CMG-S2-pow, HG (a=3), and CMG-S1-log methods are consistently ranked within the top 10 with small ARRMSE. In contrast, the CMG-S1-n, HGN (a=3), HGN (a=4), ZS, UIP and Flat methods are ranked within the bottom 10 in terms of larger ARRMSE. As expected, the Flat method is the worst across these cases. Moreover, as the between-trial variability decreases (from top to bottom panels) or the magnitude of moderation effect increases (from right to left), the distinctions among methods and the advantages of CMG methods become less evident. Specifically, the setting of high between-trial variability and weak effect moderation (see top right panel) presents the most substantial differences among methods. In this setting, the best CMG prior method, CMG-S3-log, reduces ARRMSE around 50\% compared to the Flat, HS, and SSVS methods, consolidating the impacts of magnitude of moderation effect in Figure \ref{simu:ht}.

\begin{figure}[t]
	\centering
	\includegraphics[scale=0.5]{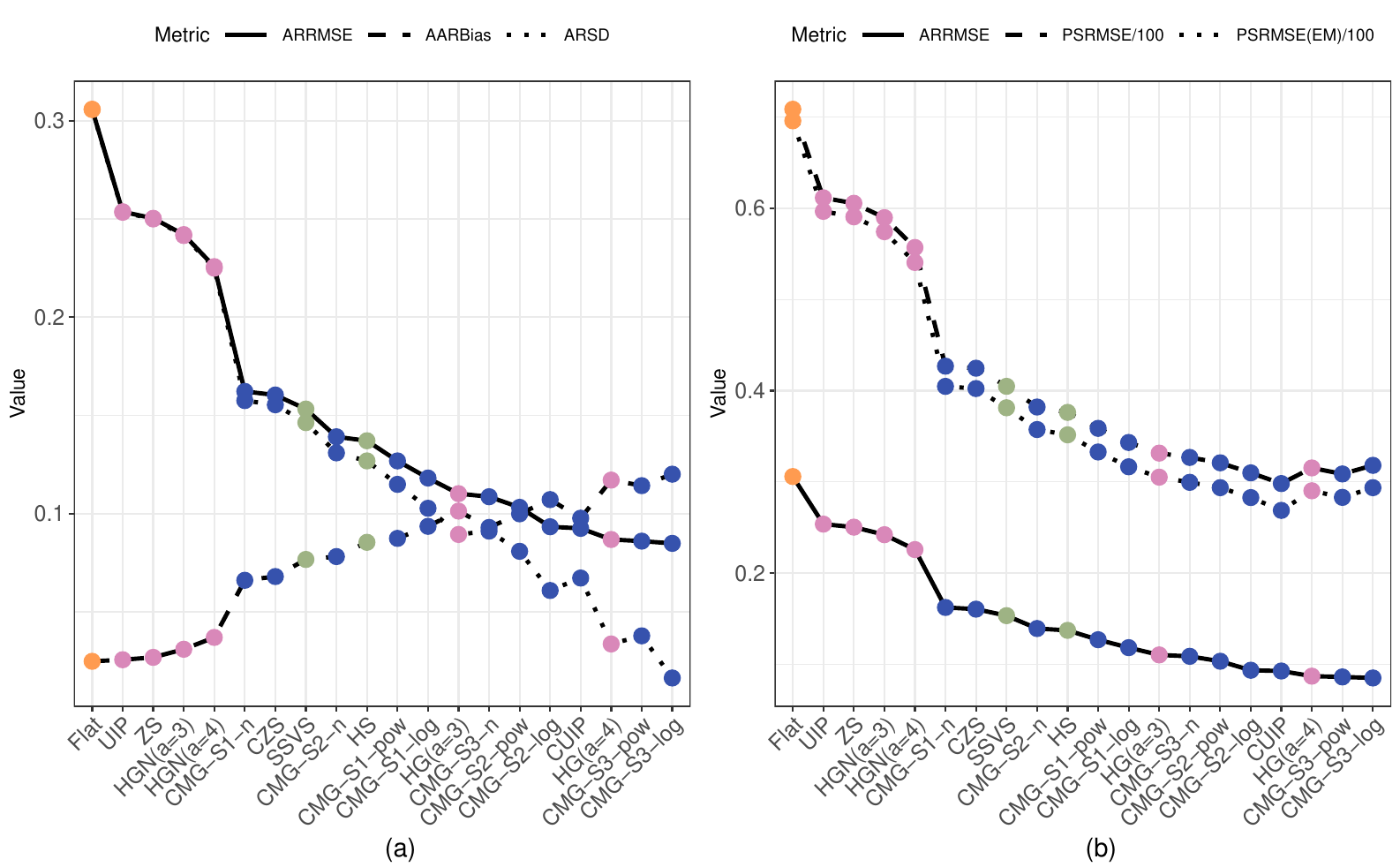}
	\caption{Simulation results under the setting of high model sparsity, high between-trial variability, weak effect moderator, and correlated covariates. Panal (a) shows ARRMSE (solid), ARSD (dotted), and AARBias (dashed) of moderation effects ($\bfgamma$) across 20 methods with a descending order of ARRMSE from left to right. Panel (b) shows ARRMSE (solid) of $\bfgamma$, PSRMSE/100 (dashed) of $\alpha$ and $\bfgamma$, and PSRMSE(EM)/100 (dotted) of $\bfgamma$ across 20 methods with a descending order of ARRMSE from left to right. The blue, pink, green and orange dots
		correspond to the CMG, NMG, HS and SSVS, and Flat methods, respectively.}
	\label{simu:trend}
\end{figure}

Figure \ref{simu:trend} compares the dynamic trends of multiple metrics across the 20 methods under the setting of the high model sparsity, weak effect moderators, correlated baseline covariates, and high between-trial variability of effect moderation. Panel (a) overlays ARRMSE, AARBias, and ARSD, and presents methods in the descending order of ARRMSE from left to right. As ARRMSE decreases, AARBias tends to increase while ARSD tends to decrease. The trend of ARRMSE is primarily driven by ARSD since the decrease in ARSD is dominant over the increase in AARBias, indicating the bias-variance trade-off. For example, comparing the CMG-S3-log method (far right) with the Flat method (far left), ARRMSE reduces more than 0.2 units at the cost of less than a 0.1 unit increase in AARBias. In addition, ARRMSE drops sharply when moving sequentially from the Flat method to CMG-S1-n method. All NMG methods (pink dots) on the left side of the CMG-S1-n prior improve ARRMSE at a negligible sacrifice in bias, since the sacrifice is so small that we can not visually distinguish ARSD and ARRMSE in the figure. In contrast, the ARRMSE decreases more slowly when shifting from the CMG-S1-n to CMG-S3-log method, accompanied by a noticeable rise in AARBias. However, the faster decline in ARSD offsets the AARBias increase, resulting in an overall reduction in ARRMSE.

Panel (b) in Figure \ref{simu:trend} further compares the parameter-level estimation (measured by ARRMSE), and participant-level estimation (measured by PSRMSE and PSRMSE(EM)), and present methods in the descending order of ARRMSE from left to right. To visualize the trends together on a reasonable scale, we plot PSRMSE/100 and PSRMSE(EM)/100 alongside ARRMSE. Overall, the relative performance of methods in the participant-level estimation aligns with their parameter-level performance. However, PSRMSE and PSRMSE(EM) exhibit sharper reductions compared to ARRMSE, highlighting the greater benefits of using the CMG methods for participant-level estimation. This also illustrates how a small improvement in parameter-level risk can translate into a larger impact in participant-level metrics. Notably, the differences between PSRMSE and PSRMSE(EM) are minimal for the Flat, UIP, ZS, and HGN methods, but become more pronounced for the other methods. This widening difference reflects more effective shrinkage in the moderation effects, accounting for participant-level characteristics. 

\section{Data analysis} \label{realdata}

\subsection{Major depression disorder dataset}
\subsubsection{Data description}
We demonstrate our methods using a real IPD-MA data example of four randomized controlled trials (RCTs) comparing efficacy between duloxetine and vortioxetine for patients having major depressive disorder (MDD) ~\citep{mdd2013,mdd2015,mdd2012,mdd2014}. Detailed description of these trials can be found elsewhere~\citep{carly2023}.

The primary outcome is the change in Montgomery-$\mathring{A}$ Depression Rating Scale (MADRS) total score~\citep{madrs1979} from baseline to the last observed follow-up. A higher MADRS score indicates a more severe symptoms. Twelve baseline covariates are considered as potential effect moderators: age, sex (female or male), smoking status (ever smoked or never smoked), weight, baseline MADRS score, baseline Hamilton Anxiety Rating (HAM-A) score~\citep{hama1959}, three comorbidity indicators for diabetes mellitus (DM), hypothyroidism, and anxiety, and three medication indicators for antidepressant, antipsychotic, and thyroid medication. For simplicity, we excluded 3 participants having incomplete baseline covariates, resulting in IPD-MA with a total of 1,844 participant.

\subsubsection{Model fitting and comparison details}

The goal of this data analysis is to assess meaningful effect moderators that alter treatment effects between duloxetine and vortioxetine. We fit the IPD-MA models in \eqref{model} with and without between-trial variability for moderation effects (i.e., with and without $\bfu_i$), using 20 methods listed in Table \ref{summary:methods}. To evaluate whether the inclusion of between-trial variability improves the model, we use the deviance information criterion (DIC)~\citep{dic2002}, with smaller DIC values indicating better model fit.

The IPD-MA models without between-trial variability of moderation effects fit our data slightly better than those with between-trial variability, as indicated by smaller DIC values across all methods. Given the typically small number of moderators in practice, in addition to the commonly used Flat method, we select the HG (a = 4) and CMG-S3-pow methods based on the best approaches identified through comparable simulation settings, as shown in the two lower panels of Figure \ref{simu:pc}. 

We then utilize the three selected methods to illustrate the application of the proposed methods in assessing effect moderation. Specifically, we focus on the posterior standard deviation (SD), 95\% credible intervals (CI), and the marginal posterior density plots of the covariate-by-treatment interaction term $\gamma_k$ in \eqref{model}. We check whether the 95\% CI excludes 0 to identify important effect moderators. However, the 95\% CI is often overly wide, potentially excluding meaningful effect moderators. This may occur when the sample size is not large enough or there is high variability in the data, which is very common in multiple studies. As a complement measure, we calculate posterior probabilities of $\gamma_k$ falling within the interval $(-\sqrt{VAR(\gamma_k|\bfy)},\sqrt{VAR(\gamma_k|\bfy)})$. A larger posterior probability of being in this interval indicates a weaker or less important effect moderator. Furthermore, the identification of important effect moderator based on such posterior probability with a threshold is known as the scaled neighborhood criterion ~\citep{li2010}. 

Accordingly, for each method, the scaled neighborhood criterion may identify a different set of important moderators with conclusions varying based on the chosen cutoff. In the absence of a specific guideline or prior information, we adopt a threshold of 0.5, following the recommendation of \cite{li2010}. Specifically, if $p_{\gamma_k}=P\left(|\gamma_k| < \sqrt{VAR(\gamma_k|\bfy)}|\bfy \right)  < 0.5 $, we consider $\gamma_k$ as an important moderation effect. This implies that there is a less than 50\% chance that the moderator lies near zero conditional on this dataset, and thus indicates the moderator is more likely to have a meaningful impact. Compared to the 95\% CI, the scaled neighborhood criterion typically yields narrower intervals. Simulations by ~\cite{li2010} also show that, while 95\% CI tends to deliver higher sensitivity, the scaled neighborhood criterion is more likely to outperform in specificity, under comparable conditions.

\subsection{Data analysis results}

\begin{figure}[t]
	\centering
	\includegraphics[scale=0.65]{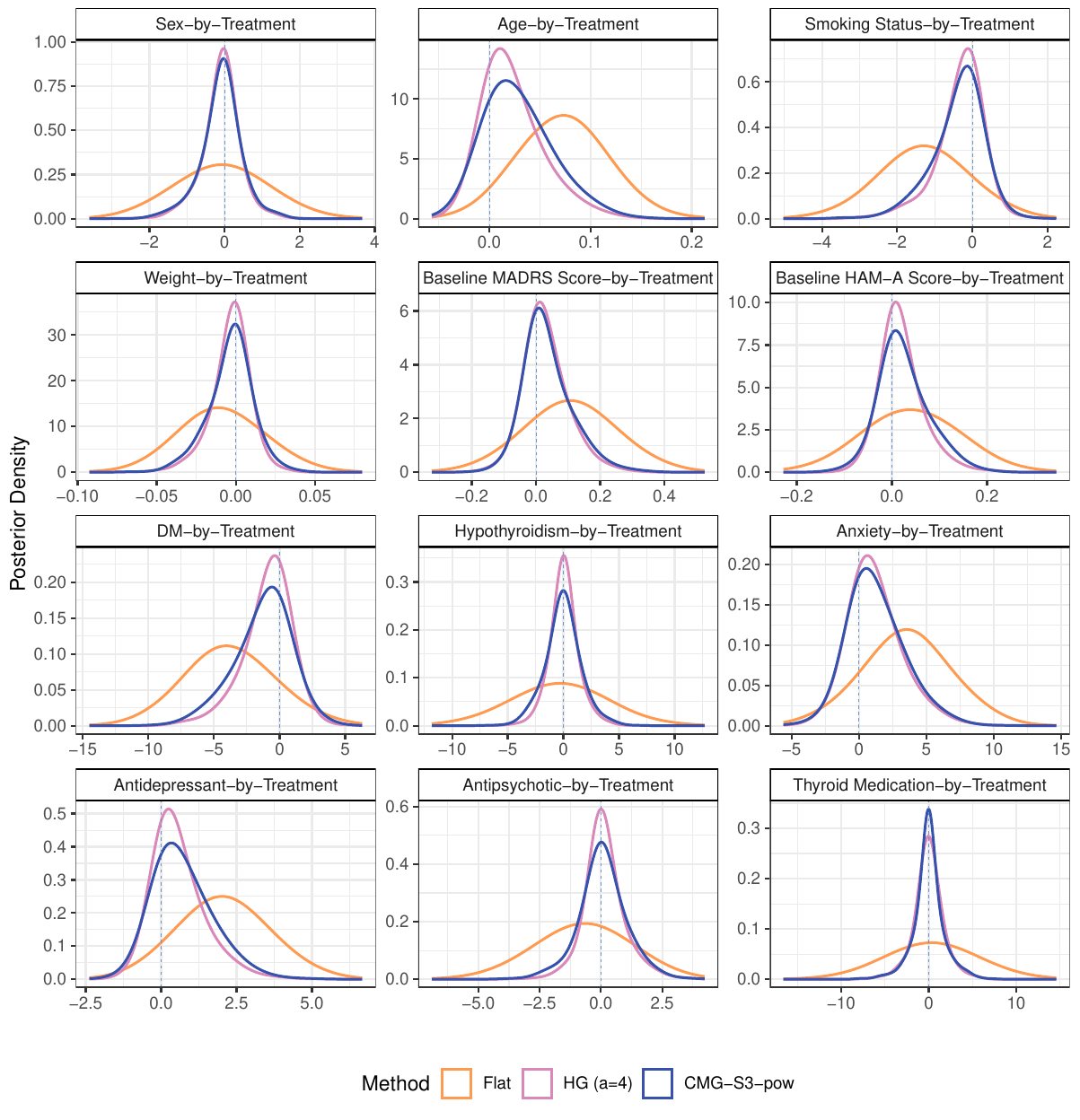}
	\caption{Posterior density plots for moderation effects (covariate-by-treatment interactions $\gamma_1,\cdots,\gamma_{12}$) under the HG (a=4), CMG-S3-pow, and Flat methods from the IPD-MA real data analysis.}
	\label{realdata:em:density}
\end{figure}

Table \ref{realdata:point:estimate} displays the posterior means, SDs and 95\% CIs for key model coefficients, including the conditional treatment effect ($\alpha$), moderation effects (covariate-by-treatment interaction $\bfgamma$), and between-trial variability of the treatment effect ($\tau_{\alpha}^2$), estimated using the Flat, HG (a=4), and CMG-S3-pow methods. The conditional treatment effect estimates are consistent across these methods, with 95\% CIs excluding 0, and the HG (a=4) and CMG-S3-pow methods producing slightly narrower CIs. For the moderation effects, the three methods agree on the directions for most moderators, except for hypothyroidism ($\gamma_8$) and antipsychotic ($\gamma_{11}$). Furthermore, while all moderation effect estimates have 95\% CIs that include 0 across the three methods, both HG (a=4) and CMG-S3-pow methods greatly improve their precisions, reducing posterior SDs by about 50\%. The HG(a=4) and CMG-S3-pow methods perform similarly, where HG(a=4) shows slightly smaller SDs but CMG-S3-pow is favored by smaller DIC.  
\begin{table}[t]
	\centering
	\caption{Posterior means (posterior SDs) and 95\% CIs of selected model coefficients ($\alpha$, $\bfgamma$ and $\tau_{\alpha}^2$) estimated using the Flat, HG (a=4), and CMG-S3-pow methods from the IPD-MA real data analysis.}
	\vspace{1em}
	\label{realdata:point:estimate}
	\resizebox{0.78\linewidth}{!}{%
		\begin{tabular}{lcccc}
			\specialrule{.1em}{.05em}{.05em}
			\textbf{Selected Coefficients}  & \textbf{Flat}  &\textbf{HG (a=4)}    & \textbf{CMG-S3-pow}   \\ \specialrule{.1em}{.05em}{.05em}
			\textbf{$\alpha$: Conditional Treatment Effect }              &2.64 (1.08)& 2.55 (0.85)      & 2.56 (0.93)         \\
			\textbf{}              &(0.44, 4.63)& (0.93, 4.20)      & (0.82, 4.29)         \\ \hline
			\textbf{$\gamma_{1}$: Sex-by-Treatment }              &-0.04 (1.08)& -0.08 (0.49)      & -0.08 (0.53)         \\ 
			\textbf{}                  & (-2.12, 2.05)& (-1.19, 0.91)     & (-1.35, 1.02)       \\ \hline
			\textbf{$\gamma_{2}$: Age-by-Treatment }            &0.07 (0.04)& 0.02 (0.03)       & 0.03 (0.03)           \\ 
			\textbf{}                 & (-0.003, 0.15) & (-0.01, 0.09)     & (-0.02, 0.1)          \\ \hline
			\textbf{$\gamma_{3}$: Smoking Status-by-Treatment }        &-1.28 (1.07) & -0.34 (0.63)      & -0.39 (0.67)         \\ 
			\textbf{}                &(-3.42, 0.78)   & (-1.97, 0.61)     & (-1.93, 0.66)       \\ \hline
			\textbf{$\gamma_{4}$: Weight-by-Treatment }       &-0.01 (0.02)  & -0.003 (0.01)       & -0.003 (0.01)         \\ 
			\textbf{}                 &(-0.06, 0.04)  & (-0.03, 0.02)     & (-0.03, 0.02)       \\ \hline
			\textbf{$\gamma_{5}$: Baseline MADRS Score-by-Treatment }    &0.11 (0.13) & 0.03 (0.07)       & 0.04 (0.07)          \\ 
			\textbf{}                   &(-0.14, 0.37)& (-0.08, 0.20)     & (-0.09, 0.21)      \\ \hline
			\textbf{$\gamma_{6}$: Baseline HAM-A Score-by-Treatment }    &0.04 (0.09)& 0.02 (0.05)       & 0.02 (0.05)       \\ 
			\textbf{}                  &(-0.14, 0.22) & (-0.06, 0.13)     & (-0.06, 0.14)      \\ \hline
			\textbf{$\gamma_{7}$: DM-by-Treatment }          &-3.78 (3.01)   & -0.95 (1.78)      & -1.24 (2.03)     \\ 
			\textbf{}                   &(-9.56, 2.36)& (-5.26, 1.96)     & (-6.07, 2.09)      \\ \hline
			\textbf{$\gamma_{8}$: Hypothyroidism-by-Treatment }   & -0.45 (3.92)  & 0.03 (1.27)       & -0.002 (1.48)        \\ 
			\textbf{}                  &(-8.14, 7.36) & (-2.47, 2.77)     & (-3.12, 3.2)        \\ \hline
			\textbf{$\gamma_{9}$: Anxiety-by-Treatment }       &3.54 (2.88)   & 1.15 (1.86)       & 1.25 (1.97)        \\ 
			\textbf{}                   &(-2.17, 9.19)  & (-1.93, 5.60)     & (-1.86, 5.91)       \\ \hline
			\textbf{$\gamma_{10}$: Antidepressant-by-Treatment } &2.04 (1.33)& 0.59 (0.83)       & 0.69 (0.89)        \\ 
			\textbf{}                  &(-0.61, 4.7) & (-0.58, 2.64)     & (-0.61, 2.76)       \\ \hline
			\textbf{$\gamma_{11}$: Antipsychotic-by-Treatment }     &-0.73 (1.72) & 0.10 (0.80)       & 0.08 (0.9)         \\ 
			\textbf{}                 &(-4.11, 2.61)  & (-1.52, 1.94)     & (-1.95, 2.06)      \\ \hline
			\textbf{$\gamma_{12}$: Thyroid Medication-by-Treatment }  &  0.51 (4.85)    & 0.05 (1.62)       & 0.06 (1.77)         \\ 
			\textbf{}                 & (-8.91, 9.76) & (-3.46, 3.41)     & (-3.74, 4.05)      \\ \hline
			\textbf{$\tau_{\alpha}^2$: Between-trial Variability of Treatment}      & 1.34 (1.01)      & 1.06 (0.83)       & 1.09 (0.89)         \\ 
			\textbf{}                & (0.07, 3.86)  & (0.05, 3.10)      & (0.07, 3.32)       \\ \specialrule{.1em}{.05em}{.05em}
		\end{tabular}%
	}
\end{table}
Figure \ref{realdata:em:density} shows the marginal posterior distributions of $\gamma_k$ under the Flat (orange lines), HG (a=4) (pink lines) and CMG-S3-pow (blue lines) methods. We assess the importance of each effect moderators by visually comparing the concentration and deviation from the null across methods. If the posterior density of $\gamma_k$ remains shifted from zero under strong shrinkage (i.e., CMGS3-pow or HG (a=4) methods), it suggests that the evidence from the data is strong enough to overcome the shrinkage, and thus indicates a potentially important moderator. This is reinforced if $\gamma_k$ from the Flat method is concentrated at a non-null value, indicating an optimistic and important signal. Conversely, if $\gamma_k$ is centered around zero in both the strong shrinkage and Flat methods, as the posterior aligns with the prior's expectation of no effect. For example, for the coefficient of sex-by-treatment, the posterior densities of three methods are perfectly concentrated around zero, indicating that sex is not an important effect moderator. In contrast, for the coefficient of age-by-treatment, its posterior densities vary greatly across methods, with the Flat method showing the widest distribution and peaking above the null. The HG (a=4) and CMG-S3-pow methods produce narrower posterior distributions, concentrated at a smaller value compared to that under the Flat method. This suggests evidence of moderation effect even under strong regularization, indicating that age is a relatively important moderator. Similarly, smoking status, DM, anxiety, and antidepressant are likely to be important moderators. It is straightforward that sex, hypothyroidism, and thyroid medication are not considered as important moderators as all three methods are symmetrically centered around the null. Finally, while the Flat method shows concentrations for weight, baseline MADRS score, baseline HAM-A score, and antipsychotic that are away from zero, both the HG (a=4) and CMG-S3-pow methods are strongly concentrated around zero, suggesting little evidence for the importance of these moderators. 
Finally, Table \ref{post:prob} presents posterior probabilities of coefficients $p_{\gamma_k}$ for all covariate-by-treatment interactions under the Flat, HG(a=4) and CMG-S3-pow methods. To identify important moderators, we compare the posterior probability with a threshold such as 0.5. For example, for the sex-by-treatment interaction ($\gamma_1$), the Flat ($p_{\gamma_1}=0.68$), HG (a=4) ($p_{\gamma_1}=0.76$), and CMG-S3-pow methods do not identify sex as an important effect moderator, since the corresponding posterior probabilities are greater than 0.5. Similarly, the Flat method identify Age, DM, Anxiety, and Antidepressant as important moderators, while HG (a=4) and CMG-S3-pow do not identify any important moderators. This implies the strong shrinkage effects of the HG (a=4) and CMG-S3-pow methods, where HG (a=4) method consistently shows higher posterior probabilities compared to the CMG-S3-pow method. To assess moderators under various shrinkage levels, across 20 methods, 14 methods identify age as an important effect moderator, 8 methods identify antidepressant use, and 6 methods identify smoker, DM, and anxiety as important moderators. The remaining covariate-by-treatment interactions are not considered by any method. These results align with the conclusions drawn in Table \ref{realdata:point:estimate} and Figure \ref{realdata:em:density}. 
\begin{table}[t]
	\centering
	\caption{The posterior probability $p_{\gamma_k}$ for covariate-by-treatment interaction ($\gamma_1, \cdots, \gamma_{12}$) from the IPD-MA real data analysis.}
	\vspace{1em}
	\label{post:prob}
	\resizebox{0.78\linewidth}{!}{%
		\begin{tabular}{lcccc}
			\specialrule{.1em}{.05em}{.05em}
			\textbf{Covariate-by-Treatment Coefficients}  & \textbf{Flat}  &\textbf{HG (a=4)}    & \textbf{CMG-S3-pow}   \\ \specialrule{.1em}{.05em}{.05em}
			\textbf{$\gamma_{1}$: Sex-by-Treatment }              &0.68& 0.76      & 0.75         \\ 
			\textbf{$\gamma_{2}$: Age-by-Treatment }            &0.20 & 0.63       & 0.58           \\ 
			\textbf{$\gamma_{3}$: Smoking Status-by-Treatment }        &0.58 & 0.74      & 0.70         \\ 
			\textbf{$\gamma_{4}$: Weight-by-Treatment }       &0.66  & 0.74       & 0.73         \\ 
			\textbf{$\gamma_{5}$: Baseline MADRS Score-by-Treatment }    &0.53 & 0.73       & 0.71         \\ 
			\textbf{$\gamma_{6}$: Baseline HAM-A Score-by-Treatment }    &0.64&0.74       &0.72       \\ 
			\textbf{$\gamma_{7}$: DM-by-Treatment }          &0.38   & 0.71     & 0.69    \\ 
			\textbf{$\gamma_{8}$: Hypothyroidism-by-Treatment }   & 0.68 & 0.76       & 0.74        \\ 
			\textbf{$\gamma_{9}$: Anxiety-by-Treatment }       &0.40   & 0.68       & 0.67        \\ 
			\textbf{$\gamma_{10}$: Antidepressant-by-Treatment } &0.30& 0.69       & 0.64      \\ 
			\textbf{$\gamma_{11}$: Antipsychotic-by-Treatment }     &0.65 & 0.77       & 0.74       \\ 
			\textbf{$\gamma_{12}$: Thyroid Medication-by-Treatment }  &  0.68    & 0.76       & 0.77         \\  \specialrule{.1em}{.05em}{.05em}
		\end{tabular}%
	}
\end{table}

\section{Discussion} \label{discussion}

In this paper, we propose the CMG methods within one-stage IPD-MA to assess treatment effect moderation by incorporating a trial-level sample size tuning function and a moderator-level shrinkage parameter in the prior. Our simulation study provides a comprehensive between CMG, NMG, and two existing shrinkage methods. We further demonstrates the application of the proposed methods to assess moderation effects of two active treatments for major depression disorder. In practice, we recommend the researchers to examine multiple methods with different shrinkage levels along with the Flat method by inspecting posterior distributions (e.g., posterior density plots) of moderation effect to learn their importance rather than making a decision based solely on 95\% CIs. 

The CMG methods in \eqref{calib:gamma} can be extended by replacing $g_k$ with $g_{k,i}$ to leverage trial-level contribution at a more concrete level. 
\begin{align}
	\gamma_k &|\bfLambda^{\star},\bft,\bfx_{[k]} \sim N\left(0,[(\bft\circ \bfx_{[k]})'\bfLambda^{\star}(\bft\circ \bfx_{[k]})]^{-1}\right), \label{joint:gamma} \\ \nonumber
	&\bfLambda^{\star} = diag\left(\sigma_1^{-2}g_{k,1}^{-1}f(n_1)^{-1}{\bf1}_{n_1},\cdots,\sigma_I^{-2}g_{k,I}^{-1}f(n_I)^{-1}{\bf1}_{n_{I}}\right), ~g_{k,i}>0. \label{general:g} 
\end{align}
Although \eqref{joint:gamma} offers more flexibility, it requires a large sample size to compensate the increasing numbers of parameters. Our additional simulations (results not shown) indicate that the benefits gained using \eqref{joint:gamma} are minimal, and thus we recommend CMG in \eqref{calib:gamma}. Moreover, the CMG methods has broad potential applications, as it is grounded in the mixtures of g-priors, which has a substantial body of literature. For example, mixtures of g-priors share conceptual links with ridge regression~\citep{ridge2012} and power prior ~\citep{chen2006}. Over the years, multiple variants of mixtures of g-priors have been developed, including but not limited to extensions to analysis of variance design~\citep{Rouder:2012,Wang:2017}, cases where parameter dimensions outnumber sample sizes~\citep{yuzo2011,ridge2012}, generalized linear regression model~\citep{yingbo:2018}, and many others ~\citep{rouder2011,fous2016,jean2023,niko2023}.  

Our simulation results conclude that the CMG methods perform as well as or better than other methods in terms of higher efficiency and lower risks (e.g., ARRMSE). These advantages are especially evident in cases with sparse models, weak effect moderation, high between-trial variability, and correlated design matrices. The CMG methods accommodate a broad range of shrinkage levels, offering a full assessment of the risk measures in estimating moderation effects, varying from the most conservative to the most optimistic method. This flexibility allows researchers to tailor the prior specification according to the desired level of shrinkage. For instance, if a researcher believes that more than 50\% shrinkage is unacceptable, one might exclude certain methods like S3 from consideration. Our results suggest that the Flat method is uniformly worse than any type of shrinkage methods, indicating that the shrinkage methods are more suitable for assessing treatment effect moderation. 

In our data analysis, it is important to note that two original trials used duloxetine as a reference medication, excluding patients with prior nonresponse. This exclusion criterion did not apply to the other two MDD trials, potentially introducing bias in the estimated conditional treatment effects of duloxetine and vortioxetine. These differences highlight the importance of incorporating study-specific parameters when constructing priors using multiple studies (e.g., the trial-level sample size tuning functions the CMG methods). We investigate multiple measures and methods to assess the importance of moderators. For example, while the 95\% CIs did not identify any important moderators with all methods, the scaled neighborhood criterion identified several potentially important ones, with age being the most frequently selected. The Flat method, the most optimistic, identified five moderators, and a method with minimal prior shrinkage (e.g, CMG-S1-n) selected two. In contrast, strong shrinkage methods (e.g., CMG-S3-log and CMG-S3-pow) identified none, reflecting the most conservative conclusions. Therefore, it is important to consider the spectrum of shrinkage levels, as they reflect varying degrees of optimism or conservatism, enabling more informed decision-making and allowing us to interpret results with an awareness of these differences rather than relying solely on a single method.

In addition, we reflect on several limitations and discuss directions for future research. First, in the CMG methods, we incorporate $f(n_i | p_i)$ to account for trial-level variation, allowing the moderator-level tuning parameter $g_k$ to more effectively identify moderators. For instance, a large estimated $g_k$ may suggest an important moderator, while a small $g_k$ could indicate a non-important one. Although $g_k$ performs well in distinguishing important moderators in studies with strong signals (results not shown), its effectiveness declines as study designs become more complex. Since $g_k$ and $f(n_i | p_i)$ jointly determine the final shrinkage effect with $g_k$ playing a dominant role, future efforts will focus on refining the CMG methods by integrating $g_k$ and $f(n_i|p_i)$ into a unified quantity to identify important moderators.

Second, in the data analysis, we complement the posterior density plots with the scaled neighborhood criterion to conveniently identify important moderators. However, this criterion relies only on posterior SDs to set the size of the neighborhood (i.e., $\sqrt{VAR(\gamma_k|\bfy)}$), ignoring the posterior means (i.e., location). Incorporating both location and scale is needed because the distributions of important moderators are characterized not only by variability but also by the extent to which their estimated effects differ from zero. Exploring a neighborhood constructed using both posterior means and standard deviations may provide a more effective way to identify moderators by considering both the magnitude and uncertainty of their estimated effects.

Last but not least, while the CMG methods show promising results in assessing moderation effects both at the parameter- and participant- level, a key challenge lies in translating these methods into actionable and personalized treatment recommendations. Specifically, the question becomes: given a new patient, which treatment - based on completed trials - will yield the most favorable health outcome? To rigorously address this, we assume that the baseline characteristics of the new patient align with those of participants from the completed trials. Consequently, we can quantify the benefit one treatment offers over the other by evaluating the predictive performance of the CMG methods. This requires extensive simulation studies to assess their robustness and reliability in guiding personalized treatment strategies.

\section*{Appendix}

Supplementary Material for ``Bayesian hierarchical models with calibrated mixtures of g-priors for assessing treatment effect moderation in meta-analysis''

		\begin{enumerate} \setlength\itemsep{0.2em}
			\item Table A1 summarizes the simulation settings in Section 3 for an easy reference.
			\item Figure A2 illustrates marginal densities $\pi_{S1}(g_k)$ across three shrinkage levels $S_1$, $S_2$, and $S_3$ in CMG methods. 
			\item Figure A3 depicts the conditional densities of $g_k/(1+g_k)$ with $b_k=1.2$ across three shrinkage levels $S_1$, $S_2$, and $S_3$ in CMG methods.
			\item Figure A4 presents three sample size tuning functions $n$, $log$, and $pow$ in CMG methods.
			\item Figure A5 compares ARSD of moderation effects ($\bfgamma$) across 20 methods under a combination of three model sparsity settings and two magnitude of effect moderation settings with high between-trial variability and correlated covariates. This figure shows the improved efficiency of the CMG methods, and validates findings from Figures 1 and 3(a).
			\item Figure A6 shows AARBias of moderation effects ($\bfgamma$) across 20 methods under a combination of three model sparsity settings and two magnitude of effect moderation settings with high between-trial variability and correlated covariates. This figure shows the sacrifice in bias using the CMG methods, but the gain in ARRMSE compensate this sacrifice, echoing findings in Figure 3(a).
			\item Figure A7 shows moderation effects ($\bfgamma$) across 20 methods under a combination of three model sparsity settings and two magnitude of effect moderation settings with high between-trial variability and uncorrelated covariates. While the main manuscript shows the results under the correlated covariates, this figure shows six scenarios under the uncorrelated covariates. The advantages of CMG methods are minimal, and the main trends remain similar to correlated covariates.   
		\end{enumerate}

% ** Acknowledgements **

\section*{Acknowledgments}
	
	This paper is based on research using data from data contributors, Takeda, and Lundbeck and, that have been made available through Vivli, Inc. Vivli has not contributed to or approved, and is not in any way responsible for, the contents of this publication.
	
	The study was funded by the Patient-Centered Outcomes Research Institute (PCORI) through PCORI Award ME-2020C321145 and the National Institute of Mental Health (NIMH) through Award R01MH126856. Disclaimer: Opinions and information in this content are those of the study authors and do not necessarily represent the views of PCORI or NIMH. Accordingly, PCORI and NIMH cannot make any guarantees with respect to the accuracy or reliability of the information and data.

%% ** The bibliograhy **
%	\bibliographystyle{ba}
%	\bibliography{literature} % place <bib-data-file> in ./bib folder 

	\bibliographystyle{agsm}
	\bibliography{literature}%

@article{dic2002,
author = {Spiegelhalter, David J. and Best, Nicola G. and Carlin, Bradley P. and Van Der Linde, Angelika},
title = {Bayesian measures of model complexity and fit},
journal = {Journal of the Royal Statistical Society: Series B (Statistical Methodology)},
volume = {64},
number = {4},
pages = {583-639},
doi = {https://doi.org/10.1111/1467-9868.00353},
year = {2002}
}

@article{gelman2006,
author = {Andrew Gelman},
title = {{Prior distributions for variance parameters in hierarchical models (comment on article by Browne and Draper)}},
volume = {1},
journal = {Bayesian Analysis},
number = {3},
publisher = {International Society for Bayesian Analysis},
pages = {515 -- 534},
year = {2006},
doi = {10.1214/06-BA117A},
}

@book{james2013,
  author = {James, Gareth and Witten, Daniela and Hastie, Trevor and Tibshirani, Robert},
  publisher = {Springer},
  title = {An Introduction to Statistical Learning: with Applications in R },
  year = 2013
}

@article{carly2023ss,
  title={Methods for Integrating Trials and Non-experimental Data to Examine Treatment Effect Heterogeneity.},
  author={Carly L Brantner and Ting-Hsuan Chang and Trang Quynh Nguyen and Hwanhee Hong and Leon Di Stefano and Elizabeth A. Stuart},
  journal={Statistical science : a review journal of the Institute of Mathematical Statistics},
  year={2023},
  volume={38 4},
  pages={
          640-654
        },
  url={https://api.semanticscholar.org/CorpusID:257220035}
}

@Article{hama1959,
author = {HAMILTON, MAX},
title = {THE ASSESSMENT OF ANXIETY STATES BY RATING},
journal = {British Journal of Medical Psychology},
volume = {32},
number = {1},
pages = {50-55},
url ={https://doi.org/10.1111/j.2044-8341.1959.tb00467.x},
eprint = {https://bpspsychub.onlinelibrary.wiley.com/doi/pdf/10.1111/j.2044-8341.1959.tb00467.x},
year = {1959}
}

@misc{carly2023,
      title={Comparison of Methods that Combine Multiple Randomized Trials to Estimate Heterogeneous Treatment Effects}, 
      author={Carly Lupton Brantner and Trang Quynh Nguyen and Tengjie Tang and Congwen Zhao and Hwanhee Hong and Elizabeth A. Stuart},
      year={2023},
      eprint={2303.16299},
      archivePrefix={arXiv},
      url= {https://doi.org/10.48550/arXiv.2303.16299},
      primaryClass={stat.ME}
}

@Article{li2010,
	author = {Qing Li and Nan Lin},
	journal = {Bayesian Analysis},
	number = {1},
	pages = {151 -- 170},
	title = {{The Bayesian elastic net}},
	volume = {5},
	year = {2010},
url="https://doi.org/10.1214/10-BA506"}

@article{madrs1979, 
title={A New Depression Scale Designed to be Sensitive to Change}, 
volume={134}, 
url={https://doi.org/10.1192/bjp.134.4.382}, 
number={4}, 
journal={British Journal of Psychiatry}, 
publisher={Cambridge University Press}, 
author={Montgomery, Stuart A. and Åsberg, Marie}, 
year={1979}, 
pages={382–389}}

@phdthesis{qiao2023,
	title={Data combining using mixtures of g-priors with application on county-level female breast cancer prevalence},
	author={Qiao Wang},
	year={2022},
url={https://doi.org/10.32469/10355/91695},
	school={University of Missouri--Columbia}
}

@misc{niko2023,
      title={Meta Analysis of Bayes Factors}, 
      author={Stavros Nikolakopoulos and Ioannis Ntzoufras},
      year={2021},
      eprint={2103.13236},
      archivePrefix={arXiv},
      url={https://doi.org/10.48550/arXiv.2103.13236},
      primaryClass={stat.ME}
}

@ARTICLE{rouder2011,
  title     = "A Bayes factor meta-analysis of Bem's {ESP} claim",
  author    = "Rouder, Jeffrey N and Morey, Richard D",
  journal   = "Psychon. Bull. Rev.",
  publisher = "Springer Science and Business Media LLC",
  volume    =  18,
  number    =  4,
  pages     = "682--689",
  month     =  aug,
  year      =  2011,
  language  = "en",
  url="https://doi.org/10.3758/s13423-011-0088-7"
}

@article{jean2023,
	author = {Jean-Michel Galharret and Anne Philippe},
	url = {https://doi.org/10.1016/j.ecosta.2021.12.009},
	issn = {2452-3062},
	journal = {Econometrics and Statistics},
	pages = {161-172},
	title = {"Bayesian analysis for mediation and moderation using g-priors"},
	volume = {27},
	year = {2023}}

@article{chen2006,
  title={The relationship between the power prior and hierarchical models},
  author={Ming-Hui Chen and Joseph G. Ibrahim},
  journal={Bayesian Analysis},
  year={2006},
  volume={1},
  pages={551-574},
  url={https://api.semanticscholar.org/CorpusID:16303249}
}

@article{ridge2012,
	author = {M. Baragatti and D. Pommeret},
	issn = {0167-9473},
	journal = {Computational Statistics \& Data Analysis},
	number = {6},
	pages = {1920-1934},
	title = {A study of variable selection using g-prior distribution with ridge parameter},
	volume = {56},
	year = {2012},
	url = {https://doi.org/10.1016/j.csda.2011.11.017}}

@ARTICLE{pans2022,
  title     = "Comparing methods for statistical inference with model uncertainty",
  author    = "Porwal, Anupreet and Raftery, Adrian E",
  journal   = "Proc. Natl. Acad. Sci. U. S. A.",
  publisher = "Proceedings of the National Academy of Sciences",
  volume    =  119,
  number    =  16,
  pages     = "e2120737119",
  month     =  apr,
  year      =  2022,
  url="https://doi.org/10.1073/pnas.2120737119",
  copyright = "https://creativecommons.org/licenses/by/4.0/"
}

@article{sebri2021,
	author = {Maamar Sebri and Hajer Dachraoui},
	url = {https://doi.org/10.1016/j.resourpol.2021.102315},
	issn = {0301-4207},
	journal = {Resources Policy},
	pages = {102315},
	title = {Natural resources and income inequality: A meta-analytic review},
	volume = {74},
	year = {2021}}

@article{wee2018,
	author = {Requia, Weeberb J. and Adams, Matthew D. and Arain, Altaf and Papatheodorou, Stefania and Koutrakis, Petros and Mahmoud, Moataz},
	doi = {10.2105/AJPH.2017.303839},
	eprint = {https://doi.org/10.2105/AJPH.2017.303839},
	journal = {American Journal of Public Health},
	note = {PMID: 29072932},
	number = {S2},
	pages = {S123-S130},
	title = {Global Association of Air Pollution and Cardiorespiratory Diseases: A Systematic Review, Meta-Analysis, and Investigation of Modifier Variables},
	url = {https://doi.org/10.2105/AJPH.2017.303839},
	volume = {108},
	year = {2018}}

@article{wang2007subgroup,
	author = {Wang, Rui and Lagakos, Stephen W. and Ware, James H. and Hunter, David J. and Drazen, Jeffrey M.},
	journal = {New England Journal of Medicine},
	number = {21},
	pages = {2189-2194},
	title = {Statistics in Medicine --- Reporting of Subgroup Analyses in Clinical Trials},
	volume = {357},
    url={https://doi.org/doi:10.1056/NEJMsr077003},
	year = {2007}}

@article{berger2014subgroup,
	author = {James O. Berger, Xiaojing Wang and Lei Shen},
	journal = {Journal of Biopharmaceutical Statistics},
	number = {1},
	pages = {110-129},
	title = {A Bayesian Approach to Subgroup Identification},
	volume = {24},
	year = {2014},
  url={https://doi.org/10.1080/10543406.2013.856026}
}

@Article{caspar2023,
author = {Van Lissa, Caspar J. and van Erp, Sara and Clapper, Eli-Boaz},
title = {Selecting relevant moderators with Bayesian regularized meta-regression},
journal = {Research Synthesis Methods},
volume = {14},
number = {2},
pages = {301-322},
year={2023},
url = {https://doi.org/10.1002/jrsm.1628}}

@article{sara2004,
	author = {Sara T. Brookes and Elise Whitely and Matthias Egger and George Davey Smith and Paul A. Mulheran and Tim J. Peters},
	journal = {Journal of Clinical Epidemiology},
	number = {3},
	pages = {229-236},
url={https://doi.org/10.1016/j.jclinepi.2003.08.009},
	title = {Subgroup analyses in randomized trials: risks of subgroup-specific analyses;: power and sample size for the interaction test},
	volume = {57},
	year = {2004}}

@article{riley2020,
author = {Riley, Richard D. and Debray, Thomas P.A. and Fisher, David and Hattle, Miriam and Marlin, Nadine and Hoogland, Jeroen and Gueyffier, Francois and Staessen, Jan A. and Wang, Jiguang and Moons, Karel G.M. and Reitsma, Johannes B. and Ensor, Joie},
title={Individual participant data meta-analysis to examine interactions between treatment effect and participant-level covariates: Statistical recommendations for conduct and planning},
journal = {Statistics in Medicine},
volume = {39},
number = {15},
pages = {2115-2137},
url = {https://doi.org/10.1002/sim.8516},
year = {2020}
}

@article{hwanhee2015,
author = {Hong, Hwanhee and Fu, Haoda and Price, Karen L. and Carlin, Bradley P.},
title = {Incorporation of individual-patient data in network meta-analysis for multiple continuous endpoints, with application to diabetes treatment},
journal = {Statistics in Medicine},
volume = {34},
number = {20},
pages = {2794-2819},
url = {https://doi.org/10.1002/sim.6519},
year = {2015}
}

@article{kraemer2013,
author = {Kraemer, Helena Chmura},
title = {Discovering, comparing, and combining moderators of treatment on outcome after randomized clinical trials: a parametric approach},
journal = {Statistics in Medicine},
volume = {32},
number = {11},
pages = {1964-1973},
url = {https://doi.org/10.1002/sim.5734},
year = {2013}
}

@article{memon2019,
  title={MODERATION ANALYSIS: ISSUES AND GUIDELINES},
  author={Mumtaz Ali Memon and Jun‐Hwa Cheah and Thurasamy Ramayah and Hiram Ting and Francis Chuah and Tat Huei Cham},
  journal={Journal of Applied Structural Equation Modeling},
  year={2019},
  url={https://api.semanticscholar.org/CorpusID:226876176}
}

@article{mdd2014,
	author = {Boulenger, Jean-Philippe and Loft, Henrik and Olsen, Christina Kurre},
title={Efficacy and safety of vortioxetine (Lu AA21004), 15 and 20 mg/day: a randomized, double-blind, placebo-controlled, duloxetine-referenced study in the acute treatment of adult patients with major depressive disorder},
	isbn = {0268-1315},
	journal = {International Clinical Psychopharmacology},
	number = {3},
	volume = {29},
	year = {2014},
url={https://doi.org/10.1097/YIC.0000000000000018}
}

@article{mdd2012,
	author = {David S. Baldwin and Henrik Loft and Marianne Dragheim},
	issn = {0924-977X},
	journal = {European Neuropsychopharmacology},
	number = {7},
	pages = {482-491},
	title = {A randomised, double-blind, placebo controlled, duloxetine-referenced, fixed-dose study of three dosages of Lu AA21004 in acute treatment of major depressive disorder (MDD)},
	volume = {22},
	year = {2012},
	url = {https://doi.org/10.1016/j.euroneuro.2011.11.008}}

@article{mdd2015,
	author = {Mahableshwarkar, Atul R. and Jacobsen, Paula L. and Chen, Yinzhong and Serenko, Michael and Trivedi, Madhukar H.},
	date = {2015/06/01},
	doi = {10.1007/s00213-014-3839-0},
	isbn = {1432-2072},
	journal = {Psychopharmacology},
	number = {12},
	pages = {2061--2070},
	title = {A randomized, double-blind, duloxetine-referenced study comparing efficacy and tolerability of 2 fixed doses of vortioxetine in the acute treatment of adults with MDD},
	url = {https://doi.org/10.1007/s00213-014-3839-0},
	volume = {232},
	year = {2015}}

@article{mdd2013,
	author = {Atul R. Mahableshwarkar, Paula L. Jacobsen and Yinzhong Chen},
	eprint = {https://doi.org/10.1185/03007995.2012.761600},
	journal = {Current Medical Research and Opinion},
	note = {PMID: 23252878},
	number = {3},
	pages = {217-226},
	publisher = {Taylor & Francis},
	title = {A randomized, double-blind trial of 2.5 mg and 5 mg vortioxetine (Lu AA21004) versus placebo for 8 weeks in adults with major depressive disorder},
	volume = {29},
	year = {2013},
	url= {https://doi.org/10.1185/03007995.2012.761600}}

@article{debray2015,
  title={Get real in individual participant data (IPD) meta-analysis: a review of the methodology},
  author={Debray, Thomas PA and Moons, Karel GM and van Valkenhoef, Gert and Efthimiou, Orestis and Hummel, Noemi and Groenwold, Rolf HH and Reitsma, Johannes B and GetReal Methods Review Group},
  journal={Research synthesis methods},
  volume={6},
  number={4},
  pages={293--309},
  year={2015},
 url="https://doi.org/10.1002/jrsm.1160",
  publisher={Wiley Online Library}
}

@article{riley2004,
  title={A systematic review of molecular and biological tumor markers in neuroblastoma},
  author={Riley, Richard D and Heney, David and Jones, David R and Sutton, Alex J and Lambert, Paul C and Abrams, Keith R and Young, Bridget and Wailoo, Alan J and Burchill, Susan A},
  journal={Clinical Cancer Research},
  volume={10},
  number={1},
  pages={4--12},
  year={2004},
url={https://doi.org/10.1158/1078-0432.ccr-1051-2},
  publisher={AACR}
}

@article{kraemer2002,
  title={Mediators and moderators of treatment effects in randomized clinical trials},
  author={Kraemer, Helena Chmura and Wilson, G Terence and Fairburn, Christopher G and Agras, W Stewart},
  journal={Archives of general psychiatry},
  volume={59},
  number={10},
  pages={877--883},
  year={2002},
  publisher={American Medical Association},
url={https://doi.org/10.1001/archpsyc.59.10.877}
}

@article{ley2012,
	author = {Eduardo Ley and Mark F.J. Steel},
	url = {https://doi.org/10.1016/j.jeconom.2012.06.009},
	issn = {0304-4076},
	journal = {Journal of Econometrics},
	number = {2},
	pages = {251-266},
	title = {Mixtures of g-priors for Bayesian model averaging with economic applications},
	volume = {171},
	year = {2012}}

@article{ishwaran2005,
  title={Spike and slab variable selection: frequentist and Bayesian strategies},
  author={Ishwaran, Hemant and Rao, J Sunil},
  journal={Annals of statistics},
  pages={730--773},
  year={2005},
  url={https://doi.org/10.1214/009053604000001147},
  publisher={Institute of Mathematical Statistics}
}

@article{hara2009,
  title={A review of Bayesian variable selection methods: what, how and which},
  author={O’Hara, Robert B and Sillanp{\"a}{\"a}, Mikko J},
  journal={Bayesian analysis},
  volume={4},
  number={1},
  pages={85--117},
  year={2009},
  url={https://doi.org/10.1214/09-BA403},
  publisher={International Society for Bayesian Analysis}
}

@article{piironen2017,
  title={Sparsity information and regularization in the horseshoe and other shrinkage priors},
  author={Piironen, Juho and Vehtari, Aki},
  journal={Electronic Journal of Statistics},
  volume={11},
  number={2},
  pages={5018--5051},
  year={2017},
  publisher={The Institute of Mathematical Statistics},
  url={https://doi.org/10.1214/17-EJS1337SI}
}

@article{bl2008,
author = {Trevor Park and George Casella},
title = {The Bayesian Lasso},
journal = {Journal of the American Statistical Association},
volume = {103},
number = {482},
pages = {681-686},
year  = {2008},
publisher = {Taylor & Francis},
url = {https://doi.org/10.1198/016214508000000337},
}

@article{ssvs1993,
author = {Edward I.   George  and  Robert E.   McCulloch },
title = {Variable Selection via Gibbs Sampling},
journal = {Journal of the American Statistical Association},
volume = {88},
number = {423},
pages = {881-889},
year  = {1993},
publisher = {Taylor & Francis},
url = {https://doi.org/10.1080/01621459.1993.10476353} 
}

@article{hs2010,
	title = {The horseshoe estimator for sparse signals},
	volume = {97},
	issn = {0006-3444},
	url = {https://doi.org/10.1093/biomet/asq017},
	doi = {10.1093/biomet/asq017},
	number = {2},
	journal = {Biometrika},
	author = {Carvalho, Carlos M. and Polson, Nicholas G. and Scott, James G.},
	month = apr,
	year = {2010},
	note = {\_eprint: https://academic.oup.com/biomet/article-pdf/97/2/465/584621/asq017.pdf},
	pages = {465--480},
}

@article{sara2019,
title = {Shrinkage priors for Bayesian penalized regression},
journal = {Journal of Mathematical Psychology},
volume = {89},
pages = {31-50},
year = {2019},
issn = {0022-2496},
url = {https://doi.org/10.1016/j.jmp.2018.12.004},
author = {Sara {van Erp} and Daniel L. Oberski and Joris Mulder}
}

@article{seo2021,
author = {Seo, Michael and White, Ian R. and Furukawa, Toshi A. and Imai, Hissei and Valgimigli, Marco and Egger, Matthias and Zwahlen, Marcel and Efthimiou, Orestis},
title={Comparing methods for estimating patient-specific treatment effects in individual patient data meta-analysis},
journal = {Statistics in Medicine},
volume = {40},
number = {6},
pages = {1553-1573},
url = {https://doi.org/10.1002/sim.8859},
year = {2021}
}

@article{fous2016,
author = {Dimitris Fouskakis and Ioannis Ntzoufras},
title = {Power-Conditional-Expected Priors: Using g-Priors With Random Imaginary Data for Variable Selection},
journal = {Journal of Computational and Graphical Statistics},
volume = {25},
number = {3},
pages = {647-664},
year  = {2016},
publisher = {Taylor & Francis},
doi = {10.1080/10618600.2015.1036996},
URL = {https://doi.org/10.1080/10618600.2015.1036996},
eprint = {https://doi.org/10.1080/10618600.2015.1036996 
}

}

@article{zellner1962,
	author = { Arnold   Zellner },
	title = {An Efficient Method of Estimating Seemingly Unrelated Regressions and Tests for Aggregation Bias},
	journal = {Journal of the American Statistical Association},
	volume = {57},
	number = {298},
	pages = {348-368},
	year  = {1962},
	publisher = {Taylor \& Francis},
	url = {https://doi.org/10.1080/01621459.1962.10480664}
}

@article{yuzo2011,
	author = {Yuzo Maruyama and Edward I. George},
	title = {{Fully Bayes factors with a generalized $g$-prior}},
	volume = {39},
	journal = {The Annals of Statistics},
	number = {5},
	publisher = {Institute of Mathematical Statistics},
	pages = {2740 -- 2765},
	year = {2011},
	doi = {10.1214/11-AOS917},
	URL = {https://doi.org/10.1214/11-AOS917}
}

@article{cui2008,
	title = {{Empirical Bayes vs. fully Bayes variable selection}},
	journal = {Journal of Statistical Planning and Inference},
	volume = {138},
	number = {4},
	pages = {888-900},
	year = {2008},
	issn = {0378-3758},
	url = {https://doi.org/10.1016/j.jspi.2007.02.011},
	author = {Wen Cui and Edward I. George}
}

@article{kw1995,
	author = {Kass, Robert E. and Wasserman, Larry},
	doi = {10.1080/01621459.1995.10476592},
	issn = {1537274X},
	journal = {Journal of the American Statistical Association},
	number = {431},
	pages = {928--934},
	title = {{A reference Bayesian test for nested hypotheses and its relationship to the schwarz criterion}},
	volume = {90},
	year = {1995}
}

@article{bayarri2012,
	author = {M. J. Bayarri and J. O. Berger and A. Forte and G. Garc{\'{\i}}a-Donato},
	title = {{Criteria for Bayesian model choice with application to variable selection}},
	volume = {40},
	journal = {The Annals of Statistics},
	number = {3},
	publisher = {Institute of Mathematical Statistics},
	pages = {1550 -- 1577},
	year = {2012},
	doi = {10.1214/12-AOS1013},
	URL = {https://doi.org/10.1214/12-AOS1013}
}

@article{bayarri2019,
	author = {M. J. Bayarri and James O. Berger and Woncheol Jang and Surajit Ray and Luis R. Pericchi and Ingmar Visser},
	title = {{Prior-based Bayesian information criterion}},
	journal = {Statistical Theory and Related Fields},
	volume = {3},
	number = {1},
	pages = {2-13},
	year  = {2019},
	publisher = {Taylor & Francis},
	url = {https://doi.org/10.1080/24754269.2019.1582126}
}

@article{som:2016,
	author = {Som, Agniva and Hans, Christopher M and MacEachern, Steven N},
	title = "{A conditional Lindley paradox in Bayesian linear models}",
	journal = {Biometrika},
	volume = {103},
	number = {4},
	pages = {993-999},
	year = {2016},
	month = {10},
	doi = {10.1093/biomet/asw037},
	url = {https://doi.org/10.1093/biomet/asw037},
	eprint = {https://academic.oup.com/biomet/article-pdf/103/4/993/8339098/asw037.pdf},
}

@article{berger2014,
	author = { James  Berger  and  M. J.   Bayarri  and  L. R.   Pericchi },
	title = {{The Effective Sample Size}},
	journal = {Econometric Reviews},
	volume = {33},
	number = {1-4},
	pages = {197-217},
	year  = {2014},
	publisher = {Taylor & Francis},
	doi = {10.1080/07474938.2013.807157},
	URL = { 
	https://doi.org/10.1080/07474938.2013.807157
	},
	eprint = { 
	https://doi.org/10.1080/07474938.2013.807157
	}
}

@article{MS2016,
	author = {Min, Xiaoyi and Sun, Dongchu},
	doi = {10.1007/s10463-015-0518-9},
	issn = {1572-9052},
	journal = {Annals of the Institute of Statistical Mathematics},
	number = {4},
	pages = {877--903},
	title = {{Bayesian model selection for a linear model with grouped covariates}},
	url = {https://doi.org/10.1007/s10463-015-0518-9},
	volume = {68},
	year = {2016}
}

@book{zellner1986,
	Author = {Goel, P. K. and Zellner, Arnold},
	ISBN = {9780444877123},
	Publisher = {North-Holland Pub. Co.},
	Series = {Studies in Bayesian econometrics and statistics: v. 6},
	Title = {{Bayesian inference and decision techniques : essays in honor of Bruno de Finetti.}},
	Year = {1986}
}

@article{ZS1980,
	author = {Zellner, A. and Siow, A.},
	doi = {10.1007/BF02888369},
	issn = {00410241},
	journal = {Trabajos de Estadistica Y de Investigacion Operativa},
	number = {1},
	pages = {585--603},
	title = {{Posterior odds ratios for selected regression hypotheses}},
	url = {https://doi.org/10.1007/BF02888369},
	volume = {31},
	year = {1980}
}

@misc{som2015,
	title={{Block Hyper-$g$ Priors in Bayesian Regression}}, 
	author={Agniva Som and Christopher M. Hans and Steven N. MacEachern},
	year={2015},
	eprint={1406.6419},
	archivePrefix={arXiv},
	primaryClass={math.ST}
}

@article{yingbo:2018,
	author = {Yingbo Li and Merlise A. Clyde},
	title = {Mixtures of $g$-Priors in Generalized Linear Models},
	journal = {Journal of the American Statistical Association},
	volume = {113},
	number = {524},
	pages = {1828-1845},
	year  = {2018},
	publisher = {Taylor & Francis},
	doi = {10.1080/01621459.2018.1469992},
	URL = {https://doi.org/10.1080/01621459.2018.1469992
	}
}

@article{burke:2017,
	title = {Meta-analysis using individual participant data: one-stage and two-stage approaches, and why they may differ},
	volume = {36},
	issn = {1097-0258},
	url = {https://doi.org/10.1002/sim.7141},
	number = {5},
	journal = {Statistics in Medicine},
	author = {Burke, Danielle L. and Ensor, Joie and Riley, Richard D.},
	year = {2017},
	pages = {335--351}
}

@article{Rouder:2012,
	title = "Default Bayes factors for ANOVA designs",
	journal = "Journal of Mathematical Psychology",
	volume = "56",
	number = "5",
	pages = "356 - 374",
	year = "2012",
	issn = "0022-2496",
	url = "https://doi.org/10.1016/j.jmp.2012.08.001",
	author = "Jeffrey N. Rouder and Richard D. Morey and Paul L. Speckman and Jordan M. Province",
}

@article{Wang:2017,
	author = "Wang, Min",
	doi = "10.1214/16-BA1011",
	fjournal = "Bayesian Analysis",
	journal = "Bayesian Anal.",
	month = "06",
	number = "2",
	pages = "511--532",
	publisher = "International Society for Bayesian Analysis",
	title = "Mixtures of $g$ -Priors for Analysis of Variance Models with a Diverging Number of Parameters",
	volume = "12",
	year = "2017",
url="https://doi.org/10.1214/16-BA1011"
}

@article{Liang:2008,
	author = {Feng Liang and Rui Paulo and German Molina and Merlise A Clyde and Jim O Berger},
	title = {{Mixtures of $g$ Priors for Bayesian Variable Selection}},
	journal = {Journal of the American Statistical Association},
	volume = {103},
	number = {481},
	pages = {410-423},
	year  = {2008},
	publisher = {Taylor \& Francis},
	url = {https://doi.org/10.1198/016214507000001337}
}

	\newpage
	
\section*{Appendix}
	
	\raggedbottom
	\begin{table}[H]
		\centering
		\caption*{Table A1: Summary of simulation settings}
		\vspace{1em}
		\resizebox{\linewidth}{!}{%
			\begin{tabular}{lll}
				\specialrule{.1em}{.05em}{.05em}
				\textbf{Simulation Factors}               & \textbf{Definition of each setting}                               & \textbf{Labels in the figure}                         \\ \hline
				\textbf{Between-Trial Variability of EM} &                                                                              \\ 
				\multicolumn{3}{l}{Characterized by $(\tau_1,\cdots,\tau_k,\cdots,\tau_8)$}                                                                              \\ 
				
				Setting 1                          & High variability: $\tau_k \in$ (1.5,2.5)                                & High Between-Trial Variability of EM                           \\ 
				Setting 2                          & Medium variability: $\tau_k \in$ (0.5,1.5)                              & Medium Between-Trial Variability of EM                          \\ 
				Setting 3                          & No Variability: $\tau_k =$ 0                                        & No Between-Trial Variability of EM                              \\ \hline
				\textbf{Model Sparsity of EM}               &                       \\ 
				\multicolumn{3}{l}{Characterized by the number of non-zero coefficients in $\bfgamma=(\gamma_1,\cdots,\gamma_k,\cdots,\gamma_8)$}                              \\
				Setting 1                          & High sparsity: $\gamma_1=\gamma_2\neq0,\gamma_3=\cdots=\gamma_8=0$   & High Sparsity                         \\ 
				Setting 2                          & Medium sparsity: $\gamma_1=\gamma_2=\gamma_3=\gamma_4\neq0,\gamma_5=\cdots=\gamma_8=0$  & Medium Sparsity                         \\ 
				Setting 3                          & Low sparsity: $\gamma_1=\cdots=\gamma_6\neq0,\gamma_7=\gamma_8=0$ & Low Sparsity                          \\ \hline
				\textbf{Magnitude of EM}                    &                                      \\ 
				\multicolumn{3}{l}{Characterized by the value of ($\gamma_1,\cdots,\gamma_k,\cdots,\gamma_8$)}                                                    \\ 
				Setting 1                          & Strong EM: $\gamma_k=1.5$                                 & Strong EM                            \\ 
				Setting 2                          & Weak EM: $\gamma_k=0.75$                                  & Weak EM                            \\ \hline
				\textbf{Correlation of Baseline Covariates} &  \\ 
				\multicolumn{3}{l}{Characterized by $\rho_{ist}$ from $MVN({\bf0}, \bfSigma_i^{B})$, where  $\bfSigma_i^{B}=\sigma_{is}^{B}\sigma_{it}^{B}\rho_{ist}^{B}$ with $\sigma_{is}^{B}=\sigma_{it}^{B}=1$} \\ 
				
				Setting 1                           & High correlation: $0.5<\rho_{ist}^{B}<0.9$                 & Shown as ``correlated covariates" in the caption                 \\ 
				Setting 2                           & No correlation: $\rho_{ist}^{B}=0$                     & Shown as ``uncorrelated covariates" in the caption  \\ \specialrule{.1em}{.05em}{.05em}
			\end{tabular}%
		}
	\end{table}
	
	\begin{figure}[H]
		\centering
		\hspace{-3em}
		\includegraphics[scale=0.52]{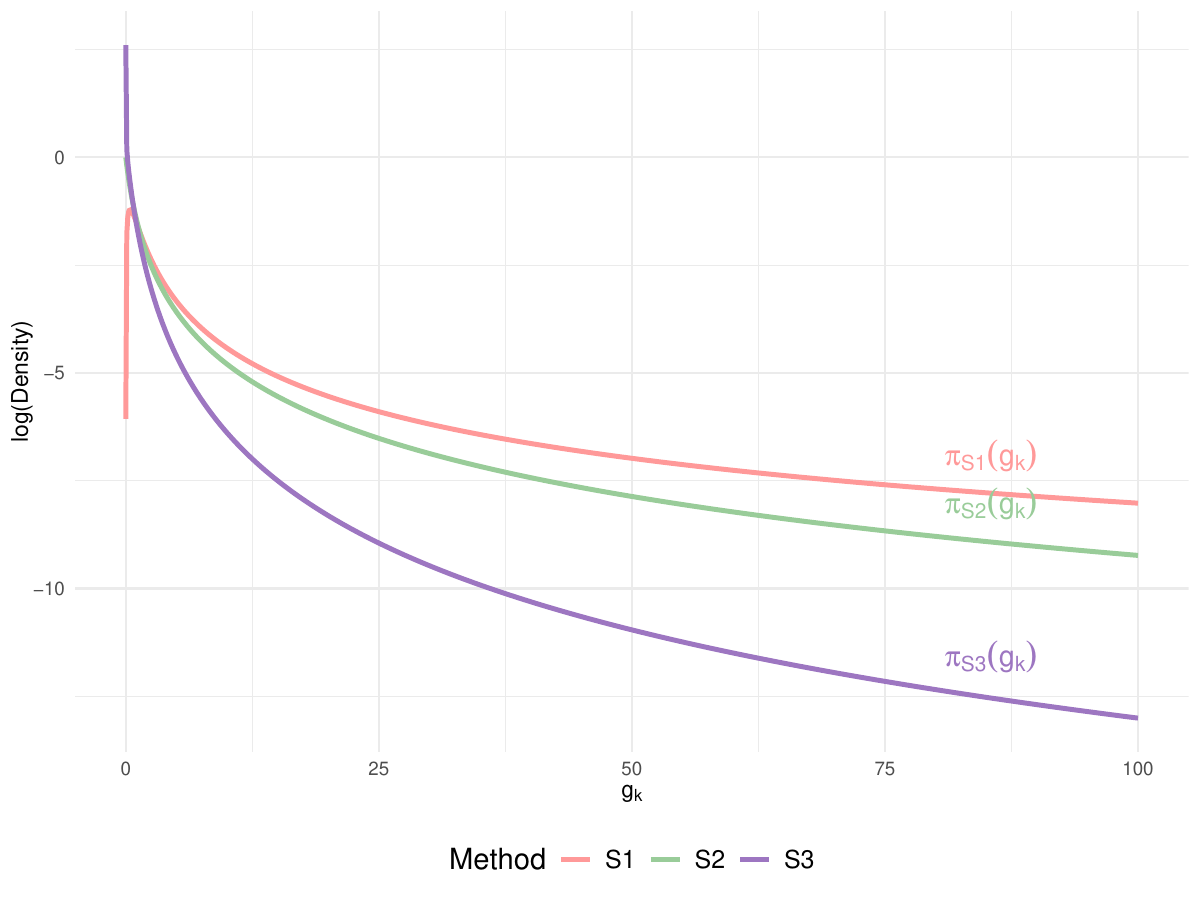}
		\caption*{\small Figure A1: Marginal density functions of $g_k$, and $\pi_{S1}(g_k)$, $\pi_{S2}(g_k)$, and $\pi_{S3}(g_k)$ (in log scale) correspond to  shrinkage levels $S1$ (red), $S2$ (green) and $S3$ (purple) in CMG methods, respectively.}
	\end{figure}
	
	\begin{figure}[H]
		\centering
		\hspace{-3em}
		\includegraphics[scale=0.52]{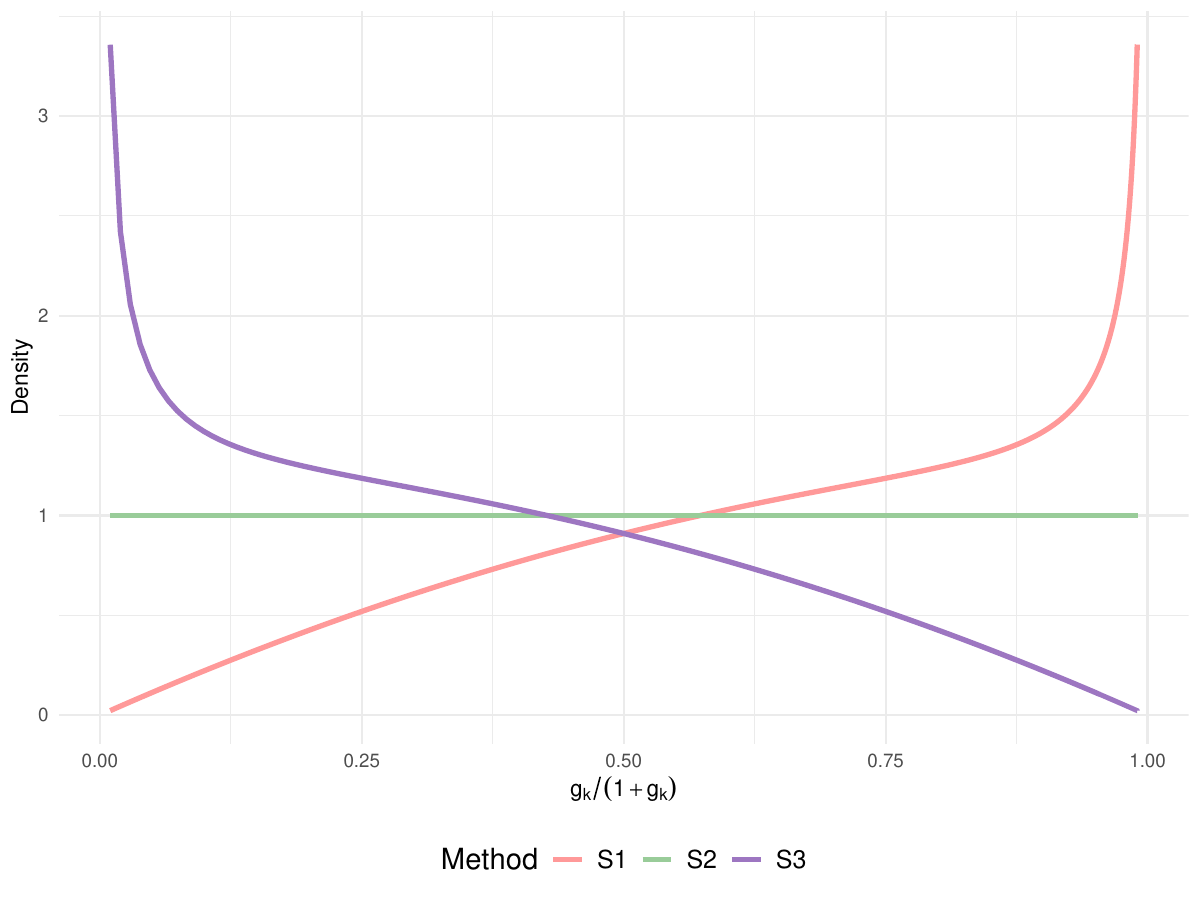}
		\caption*{\small Figure A2: Marginal density functions of $g_k/(1+g_k)$ across three shrinkage levels $S1$ (red), $S2$ (green) and $S3$ (purple) in the CMG priors.}
	\end{figure}
	
	\begin{figure}[H]
		\centering
		\hspace{-3em}
		\includegraphics[scale=0.52]{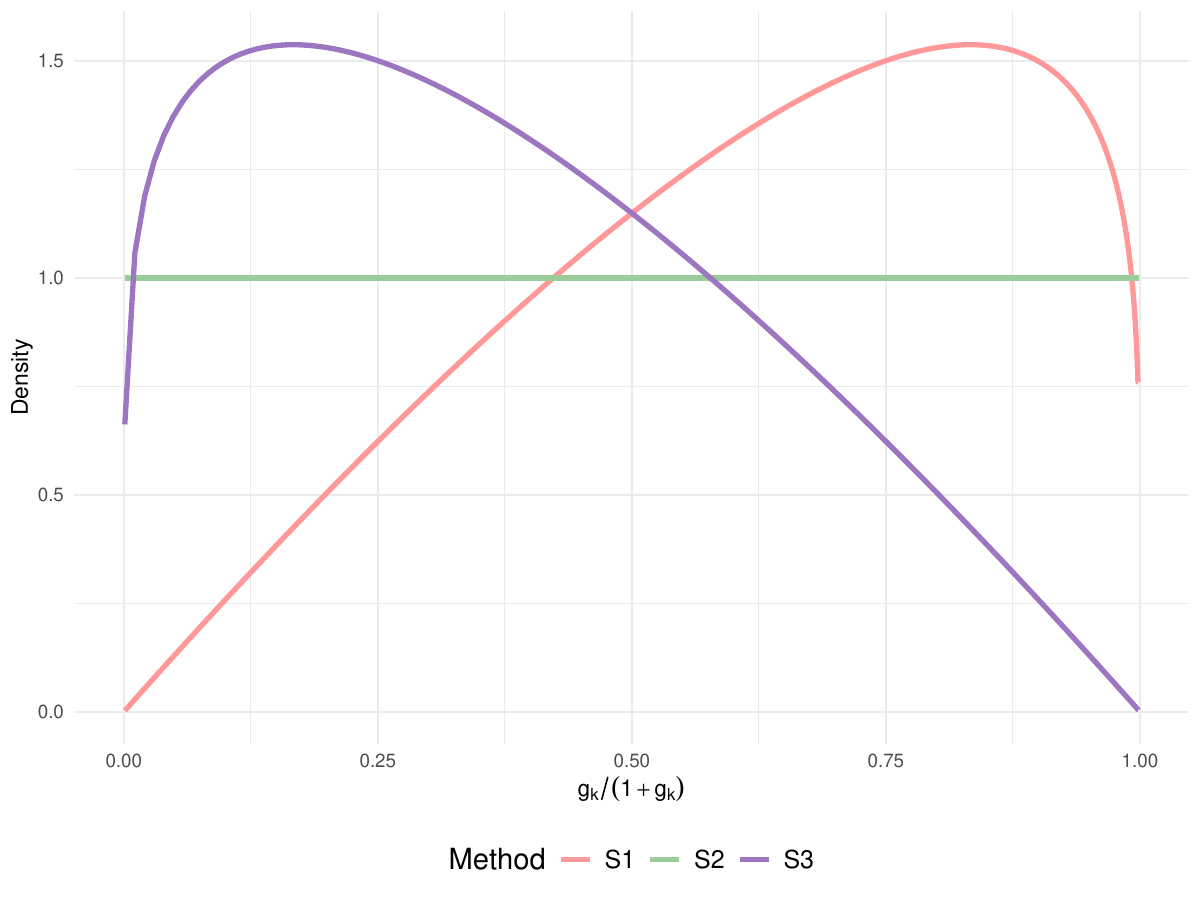}
		\caption*{\small Figure A3: Conditional density functions of $g_k/(1+g_k)$ with $b_k=1.2$ across three shrinkage levels $S1$ (red), $S2$ (green) and $S3$ (purple) in CMG methods.}
	\end{figure}

	\begin{figure}[H]
		\centering
		\hspace{-3em}
		\includegraphics[scale=0.52]{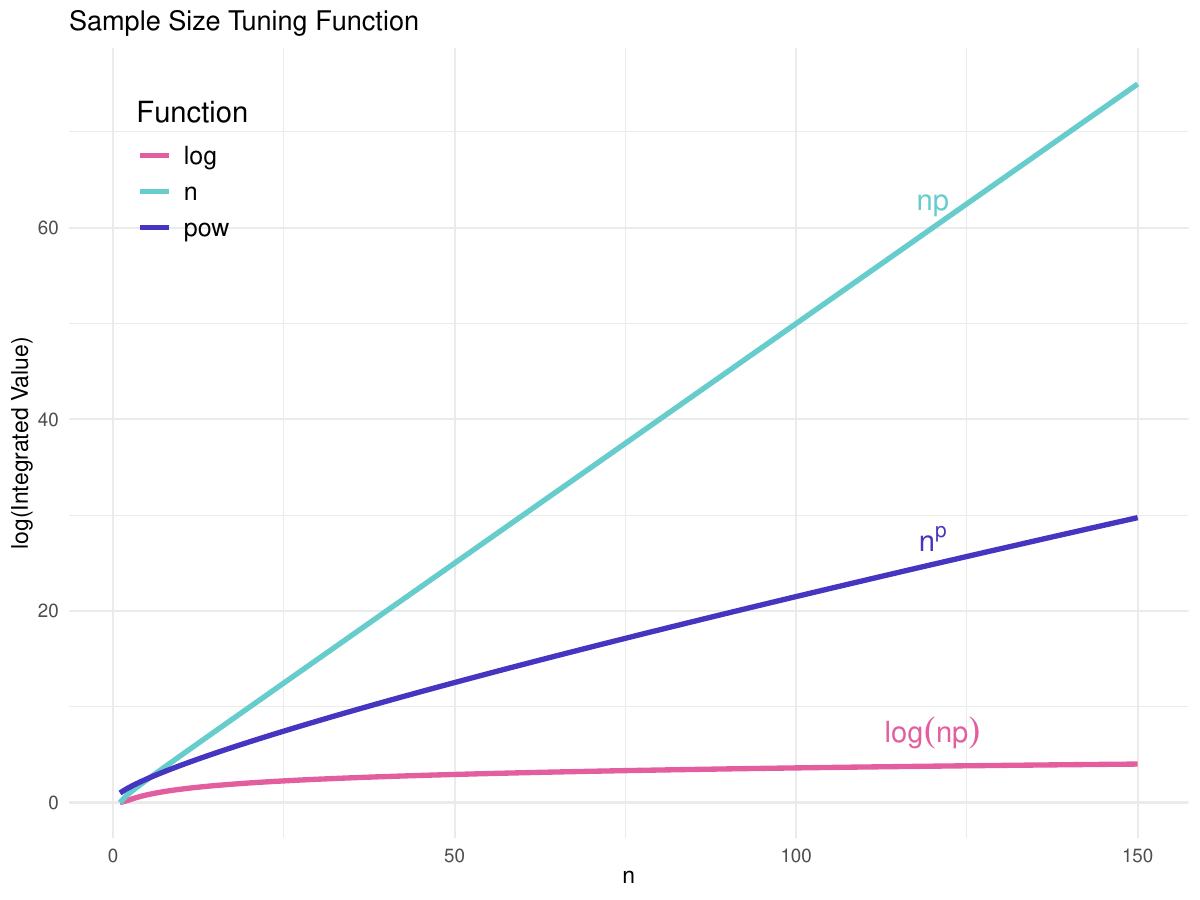}
		\caption*{\small Figure A4: Illustration of different rates of increase under three sample size tuning functions $n$ (cran), $log$ (purple) and $pow$ (pink) in CMG methods with tuning parameter $p=0.5$.}
	\end{figure}

	\begin{figure}[H]
		\centering
		\hspace{-3em}
		\includegraphics[scale=0.75]{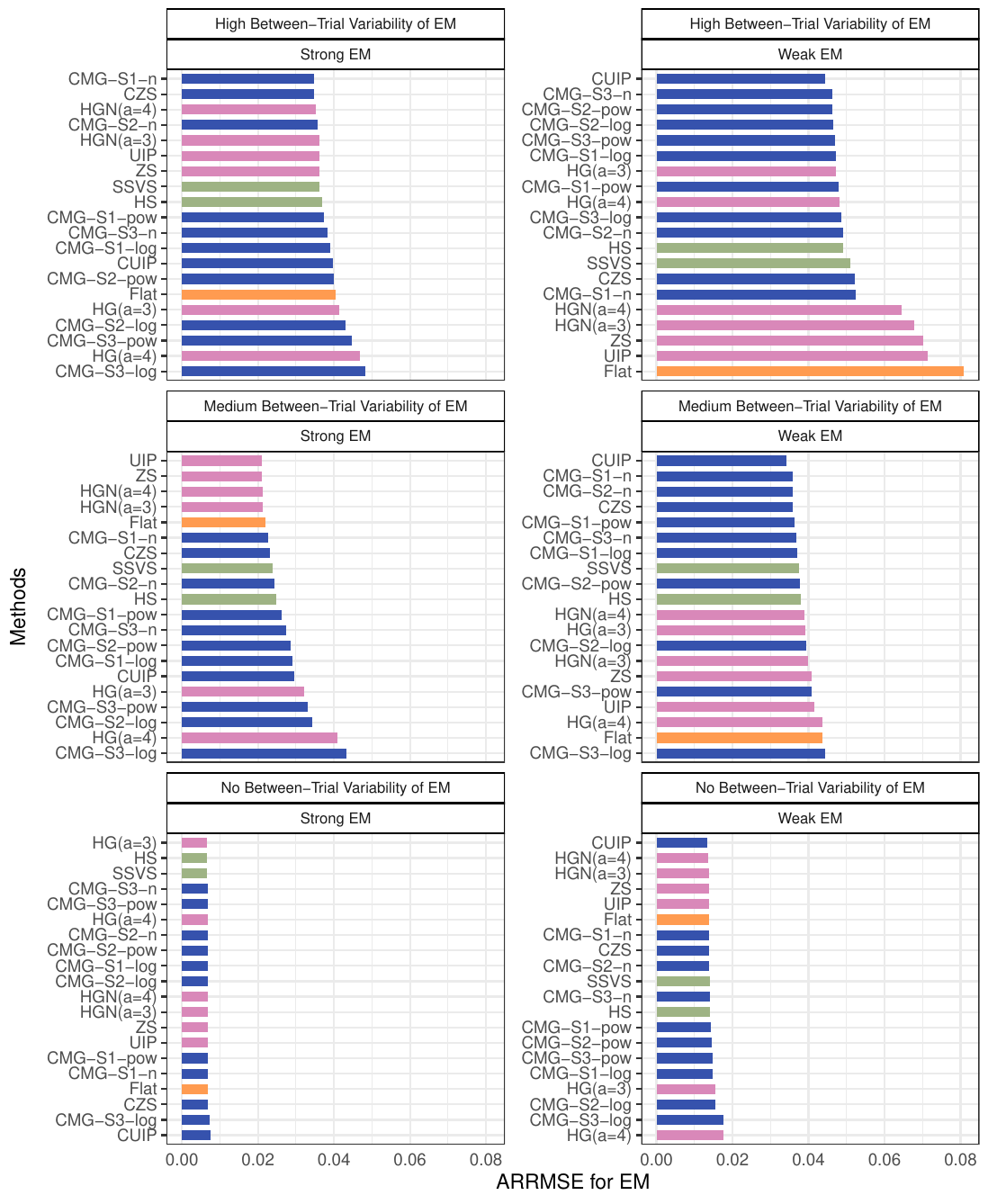}
		\caption*{\small Figure A5: ARSD of moderation effects ($\bfgamma$) across 20 methods under a combination of three model sparsity settings (from top to bottom panels) and two magnitude of effect moderation settings (left and right panels) with high between-trial variability and correlated covariates. In each panel, methods are ordered by their performance. The blue, pink, green, and orange bars correspond to the CMG, NMG, HS and SSVS, and Flat methods, respectively.}
	\end{figure}
	
	\begin{figure}[H]
		\centering
		\hspace{-3em}
		\includegraphics[scale=0.75]{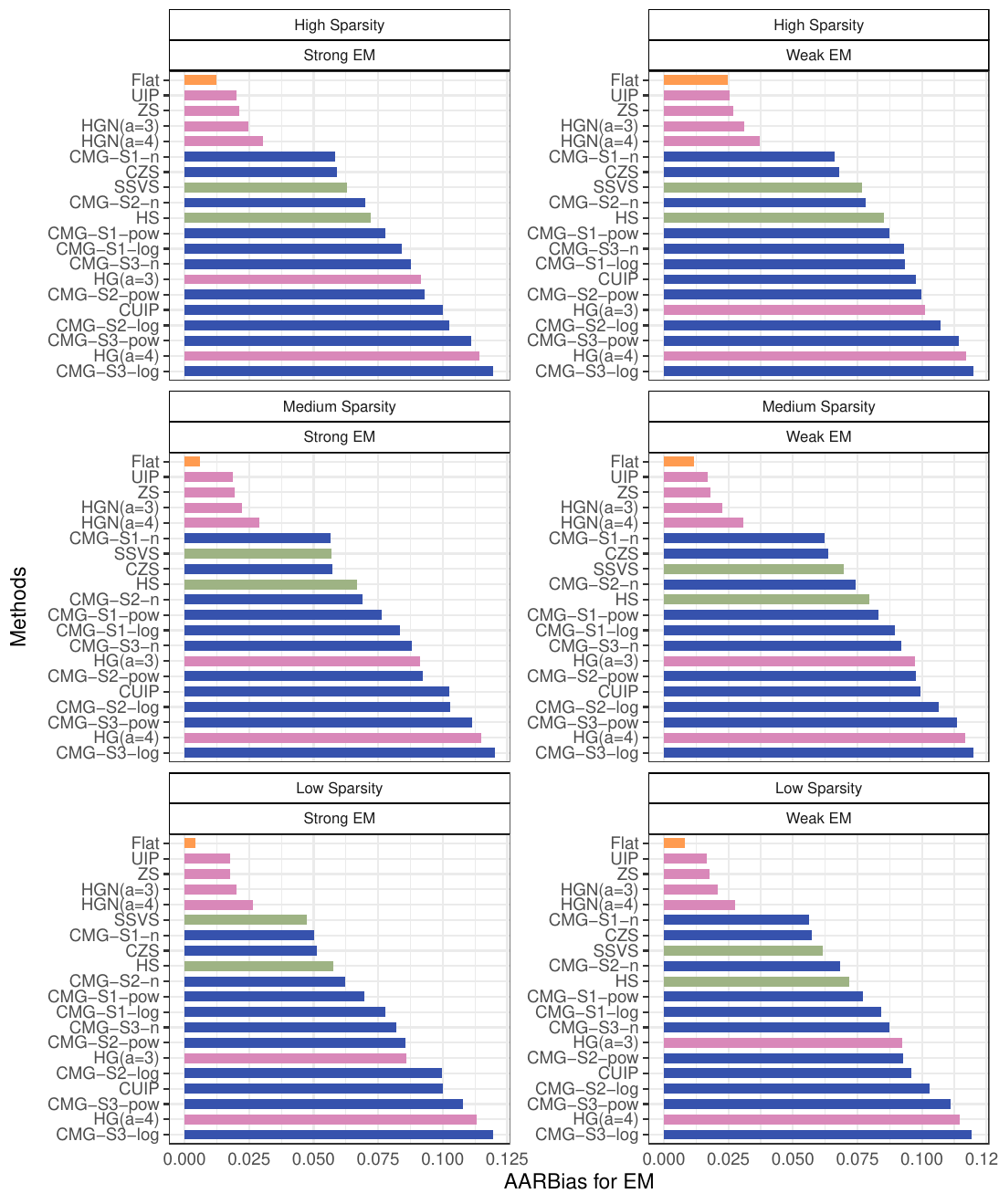}
		\caption*{\small Figure A6: AARBias of moderation effects ($\bfgamma$) across 20 methods under a combination of three model sparsity settings (from top to bottom panels) and two magnitude of effect moderation settings (left and right panels) with high between-trial variability and correlated covariates. In each panel, methods are ordered by their performance. The blue, pink, green, and orange bars correspond to the CMG, NMG, HS and SSVS, and Flat methods, respectively.}
	\end{figure}
	
	\begin{figure}[H]
		\centering
		\hspace{-3em}
		\includegraphics[scale=0.75]{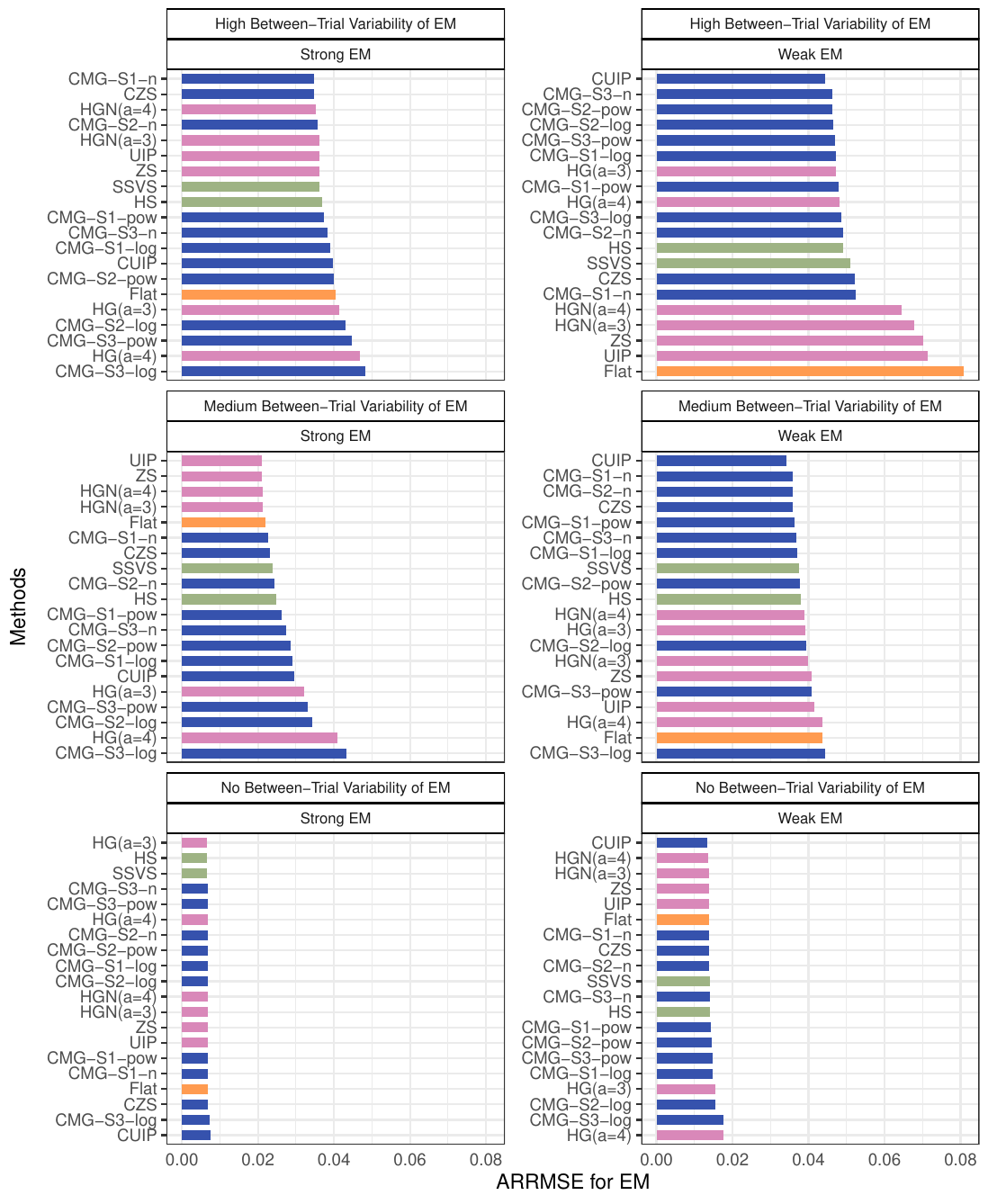}
		\caption*{\small Figure A7: ARRMSE of moderation effects ($\bfgamma$) across 20 methods under a combination of three model sparsity settings (from top to bottom panels) and two magnitude of effect moderation settings (left and right panels) with high between-trial variability and uncorrelated covariates. In each panel, methods are ordered by their performance. The blue, pink, green, and orange bars correspond to the CMG, NMG, HS and SSVS, and Flat methods, respectively.}
	\end{figure}

\end{document}